\documentclass[lettersize,journal]{IEEEtran}
\usepackage{amsmath,amsfonts}

\usepackage{algorithmic}
\usepackage{algorithm}
\usepackage{array}
\usepackage{textcomp}
\usepackage{stfloats}
\usepackage{url}
\usepackage{verbatim}
\usepackage{graphicx}
\usepackage{cite}
\usepackage{xcolor}
\usepackage{subfigure}
\usepackage{booktabs}
\usepackage{multirow}
\begin{document}

\title{Deep Learning Based Dynamic Environment Reconstruction for Vehicular ISAC Scenarios}

\author{Junzhe Song,
Ruisi He,~\IEEEmembership{Senior Member,~IEEE,}
Mi Yang,~\IEEEmembership{Member,~IEEE,}\\
Zhengyu Zhang,~\IEEEmembership{Student Member,~IEEE,}
Bingcheng Liu, Jiahui Han, Haoxiang Zhang, 
Bo Ai,~\IEEEmembership{Fellow Member,~IEEE}

\thanks{This work is supported by the National Natural Science Foundation of China under Grant 62431003 and 62271037. (Corresponding authors: Ruisi He.)}

\thanks{
J. Song, R. He, M. Yang, Z. Zhang, and B. Ai are with the State Key Laboratory of Advanced Rail Autonomous Operation, the School of Electronics and Information Engineering, and the Frontiers Science Center for Smart High-speed Railway System, Beijing Jiaotong University, Beijing 100044, China (email: 23125052@bjtu.edu.cn; ruisi.he@bjtu.edu.cn; myang@bjtu.edu.cn; 21111040@bjtu.edu.cn; boai@bitu.edu.cn).

Bingcheng Liu is with Aerospace Information Research Institute, Chinese Academy of Sciences, Beijing 100094, China (e-mail: liubc@aircas.ac.cn).

Jiahui Han and Haoxiang Zhang are with China Academy of Industrial Internet, Ministry of Industry and Information Technology, Beijing, China (email: hjh1760708@126.com; zhx61778294@126.com).

}}

\maketitle

\begin{abstract}
Integrated Sensing and Communication (ISAC) technology plays a critical role in future intelligent transportation systems, by enabling vehicles to perceive and reconstruct the surrounding environment through reuse of wireless signals, thereby reducing or even eliminating the need for additional sensors such as LiDAR or radar. However, existing ISAC-based reconstruction methods often lack the ability to track dynamic scenes with sufficient accuracy and temporal consistency, limiting the real-world applicability. To address this limitation, we propose a deep learning based framework for vehicular environment reconstruction by using ISAC channels. We first establish a joint channel–environment dataset based on multi-modal measurements from real-world urban street scenarios. Then, a multi-stage deep learning network is developed to reconstruct the environment. Specifically, a scene decoder identifies the environmental semantic context such as buildings, trees, and so on; a semantic center decoder predicts coarse spatial layouts by localizing dominant object centers; a point cloud decoder recovers fine-grained geometry and structure of surrounding environments. Experimental results demonstrate that the proposed method achieves high-quality global reconstruction of dynamic environments, with a Chamfer Distance of 0.29 and F-score@0.5m of 0.87. In addition, complexity analysis demonstrates the efficiency and practical applicability of the method in real-time scenarios. This work provides a pathway toward low-cost environment reconstruction based on ISAC for future intelligent transportation.
\end{abstract}

\begin{IEEEkeywords}
ISAC, Dynamic Environment Reconstruction,  Channel Measurement, Deep Learning.
\end{IEEEkeywords}

\IEEEdisplaynontitleabstractindextext

\IEEEpeerreviewmaketitle

\section{Introduction}
\IEEEPARstart{I}{ntegrated} Sensing and Communication (ISAC) has been identified by International Telecommunication Union Radiocommunication Sector (ITU-R) as a key feature of next-generation networks and is recognized as one of the six core visions of IMT-2030 (6G) \cite{10812728,he2024wireless,tian2025analytical}. ISAC aims to achieve a deep integration of communication and sensing functionalities by leveraging shared spectrum and hardware resources \cite{10506595,zhang2024cluster}. In ISAC-enabled scenarios, network infrastructure is no longer limited to data transmission and connectivity, it also serves as a distributed sensing platform for environmental imaging and three-dimensional (3D) reconstruction tasks. Millimeter-wave (mmWave) frequencies, with their large bandwidth, narrow beamwidth, and high spatial resolution, offer unique advantages in ISAC systems \cite{he2019propagation,he2017geometrical}. These characteristics make mmWave ISAC particularly effective for target detection, scene modeling, and environmental perception \cite{wei2023integrated}. In vehicular ISAC scenarios, vehicles can use existing communication systems to perceive the surrounding environment. This offers a low-cost, highly integrated solution that reduces the dependency on expensive dedicated sensing equipment, such as LiDAR or radar, particularly for large-scale deployment, thereby significantly lowering the cost of environmental perception \cite{10872967}. Unlike traditional static ISAC scenarios, vehicular ISAC faces unique challenges such as rapidly changing environments and continuously varying observation angles, which impose stricter requirements on perception accuracy and temporal consistency \cite{zhang2022general}. Therefore, developing dynamic environment reconstruction techniques tailored for vehicular ISAC scenarios has significant practical relevance and technical value.

Traditionally, 3D environment reconstruction using wireless signals primarily relies on Frequency-Modulated Continuous Wave (FMCW) signals transmitted by mmWave radar systems, with Fast Fourier Transform (FFT) techniques commonly employed for echo signal processing \cite{sun2020mimo}. Various work has focused on leveraging mmWave radar echoes for target detection, point cloud reconstruction, and even scene-level semantic understanding, demonstrating considerable potential in applications such as autonomous driving \cite{qiu2025multi,zifeng20254d,luo2025improving}, security and health monitoring \cite{11017419}\cite{11037950}, and Internet of Things (IoT) \cite{11028890}\cite{11015573}. However, such works typically rely on dedicated sensing hardware (e.g., LiDAR, RGB-D cameras), which are costly, power-intensive, and highly sensitive to adverse environmental conditions such as rain, fog, and lighting variations, making them unsuitable for long-term deployment in resource-constrained edge dynamic scenarios \cite{chen2025rageosense}. Moreover, sensing tasks are often designed independently and deployed redundantly, resulting in low spectral efficiency and failing to meet the demands of future ubiquitous intelligent systems for low-cost, highly integrated sensing-communication platforms \cite{chen2024enhancing}\cite{cheng2023m}. With the advancement of ISAC, communication-based environment reconstruction has become a new research trend. ISAC offers inherent advantages such as spectrum sharing, joint resource scheduling, and hardware co-design, enabling the development of highly integrated and energy-efficient 3D dynamic reconstruction systems. However, FMCW waveforms are not compatible with communication schemes, making existing reconstruction solutions difficult to adapt to ISAC applications \cite{10770127}\cite{10039194}. Therefore, the development of ISAC-based environment reconstruction methods has become critical.

In wireless communication systems, the channel refers to the physical medium through which electromagnetic signals propagate from the transmitter (Tx) to the receiver (Rx). Unlike traditional communication channels that are used exclusively for data transmission, ISAC channels simultaneously carry user signals and embed rich information of surrounding environments \cite{lu2024integrated}. Channel parameters, such as delay, azimuth angle of arrival (AoA), and elevation angle of arrival (EoA), essentially characterize the physical propagation process of signals in space, making the reconstruction of the environmental geometry from the channel information physically feasible \cite{zhang2023integrated}. Current research on ISAC-enabled environment reconstruction mainly follows two pathways: one relies on communication channel information, and the other on sensing channel feedback. The former often adopts multi-node cooperative architectures, leveraging geometric patterns and statistical variations in received signals to infer the environment. For example, Chang et al. proposed a multi-object reconstruction method by clustering reflection points along communication paths, validated under THz band experiments \cite{chang2024environment}, Lu et al. developed a communication channel-driven reconstruction framework enabling multi-user joint modeling \cite{10636720}. However, this work relies on simulation data and employs deep learning primarily for point cloud density enhancement following traditional signal processing (e.g., MUSIC, CFAR), rather than direct channel-to-geometry mapping. In contrast, reconstruction approaches based on sensing channels typically extract multi-path characteristics from the observation points to restore environmental structure. For example, Blandino et al. reconstructed UAV 3D positions based on sensing channel observations \cite{blandino2025detecting}, while Yin et al. proposed a dual-base cooperation model that utilizes uplink echoes to achieve 3D imaging \cite{10945996}. Although these studies demonstrate the feasibility of using ISAC channels for environment reconstruction, most existing works still rely on simulation data or idealized experimental setups, such as indoor scenes, static targets, or isolated object localization. Consequently, their performance degrades significantly in highly dynamic traffic scenarios. As highlighted in \cite{avazov2021trajectory}, when dynamic scatterers vary frequently, the stability and interpretability of mmWave channel parameters deteriorate, severely impacting the continuity and accuracy of 3D reconstruction. Moreover, Ref. \cite{du2022integrated} points out that current methods generally lack effective alignment mechanisms for dynamic target features, limiting their applicability in high-mobility tasks such as Vehicle-to-Everything (V2X) communication and autonomous driving. Therefore, the development of ISAC channel-based reconstruction methods adapted for dynamic environments has become urgent and critical for future research.

Although the ISAC channel inherently contains rich information about the surrounding environment, its high-dimensional and abstract features make it challenging to establish an explicit mapping to dynamic environments \cite{10644121}. In this context, deep learning excels at modeling such complex and structured data \cite{10839242,huang2022artificial}. Several studies have explored the application of deep learning techniques to tasks involving channel feature processing. For example, Nagao et al. employed a Vision Transformer architecture to process multi-channel spatiotemporal features for detecting weak seismic events \cite{katoh2025multidlformer}, Yun et al. utilized a Transformer to extract the impact of channel variations on semantic tasks, addressing multi-task optimization in dynamic wireless environments \cite{yun2025toast}; Fang et al. proposed a digital twin-assisted deep reinforcement learning (DT-DRL) framework to learn directional channel characteristics and reconstruct wireless environments \cite{11060842}, Liu et al. leveraged generative adversarial networks (GANs) to learn channel characteristics in IoV scenarios, generating adaptive channel responses to the environment \cite{liu2025ai}. These studies demonstrate the significant advantages of deep learning models in handling high-dimensional, dynamic, and structurally dependent channel characteristics. 

% Building upon this foundation, this paper develops a deep learning based dynamic environment reconstruction framework to capture the environmental information embedded in ISAC channel characteristics.

% Dynamic vehicular scenarios impose stringent requirements on 3D environment reconstruction, demanding both low-cost and high-accuracy solutions. However, traditional methods often fail to perform well in dynamic and complex environments, whereas the structural information embedded in ISAC channels opens up new possibilities for accurate and robust reconstruction. To address the aforementioned challenges, this paper proposes a dynamic 3D environment reconstruction method based on ISAC channel, integrated with deep learning. The main contributions and innovations of this work are summarized as follows:
Dynamic vehicular scenarios impose stringent requirements on 3D environment reconstruction, demanding both low-cost and high-accuracy solutions. However, most existing works often fail to perform well in dynamic and complex environments, whereas the structural information embedded in ISAC channels opens up new possibilities for accurate and robust reconstruction. Specifically, this paper utilizes ISAC channel information (multi-path parameters: delay, AoA, and EoA) for reconstruction, addressing two core challenges: 1) The difficulty of establishing an explicit mapping from the high-dimensional and abstract ISAC channel characteristics to the dynamic environment. 2) Overcoming the limitations of existing ISAC methods that rely heavily on simulation or idealized static data, demanding a robust solution for highly dynamic vehicular scenarios. To address the aforementioned challenges, this paper proposes a dynamic 3D environment reconstruction method based on ISAC channel, integrated with deep learning. The main contributions and innovations of this work are summarized as follows:

\begin{itemize}
  \item A deep learning based dynamic environment method is proposed, driven by ISAC channel data. This method exploits the physical characteristics of multi-path propagation in sensing channel to recover complete 3D environment point cloud from sparse channel parameters.
  
  \item A unified vehicle-mounted ISAC and LiDAR measurement platform is developed and deployed for synchronized data collection at 28\,GHz in real road scenarios, enabling joint acquisition of channel data and LiDAR point clouds in dynamic environments.
  
  \item A multi-stage channel-to-reconstruction network (MSCR-Net) is proposed, following a progressive training and cascade guidance strategy to sequentially achieve scene classification, semantic center localization, and point cloud generation.
\end{itemize}

The remainder of this paper is organized as follows: Section II outlines the proposed framework. Section III presents the ISAC channel measurement. Section IV details the data pre-processing. Section V details the architecture of the proposed MSCR-Net. Section VI evaluates the performance of the model. Finally, Section VII concludes the paper.

%%
%%%%%%%系统框图%%%%%%%%%%%%%
\begin{figure*}[t]
    \centering
    \includegraphics[width=\linewidth]{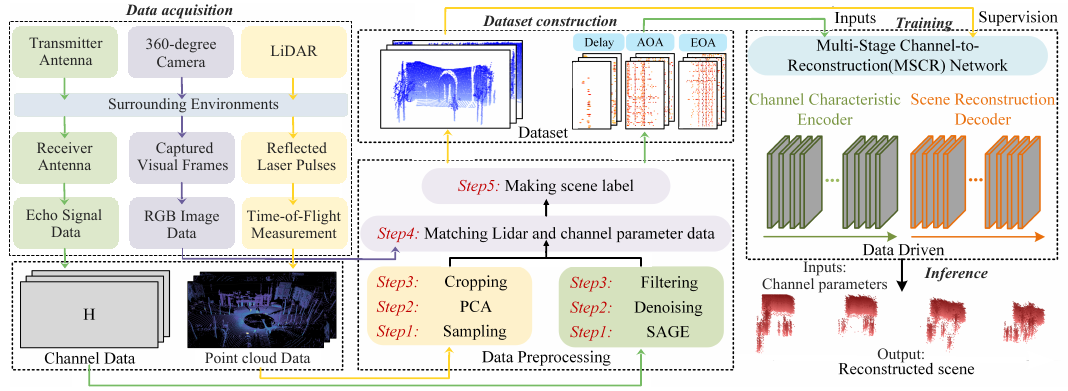}
    \caption{An overview of the proposed method framework.}
    \label{fig:systemv2}
\end{figure*}
%%%%%%%%%%%%%%%%%%%%%%%%%%
% %%%%%%%模拟场景%%%%%%%%%%%%%
% \begin{figure}[t]
%     \centering
%     \includegraphics[width=\linewidth]{模拟场景v2.pdf}
%     \caption{Layout of a typical vehicular ISAC urban scenario.}
%     \label{fig:layout}
% \end{figure}
% %%%%%%%%%%%%%%%%%%%%%%%%%%
\section{Environment Reconstruction Framework}%%%第二章
We propose a mmWave ISAC channel data driven method for dynamic 3D environment reconstruction, with the overall framework illustrated in Fig. \ref{fig:systemv2}.
The proposed method leverages both channel characteristics and LiDAR point cloud data to train a model capable of reconstructing dynamic 3D environment. ISAC channel data can typically be obtained through Ray-Tracing simulations \cite{klautau2021generating} or real-world measurements. Compared to Ray-Tracing, measurement-based data offer greater authenticity and representativeness in capturing environmental complexity and dynamic variations, making them more suitable for evaluating the generalization capability of the model. We built an integrated measurement platform, where a pair of transceivers forms the sensing link. Meanwhile, LiDAR is employed to acquire high-resolution point clouds as ground-truth supervision for model training. In addition, a RGB panoramic camera is employed to assist in identifying the scene category of each sample.

% During the data pre-processing stage, the channel data are first processed using Space Alternating Generalized Expectation Maximization (SAGE) algorithm to extract the channel parameters for each snapshot, including the propagation delay $\tau$, AOA $\theta$, and EOA $\phi$. Detailed descriptions of the SAGE algorithm can be found
% in \cite{fessler2002space}, \cite{chong2002joint}, and \cite{matthaiou2007characterization}. Subsequently, de-noising and spurious path filtering are applied to construct a set of channel feature vectors consisting of N multi-path components:
During the data pre-processing stage, the receiver first acquires the raw in-phase and quadrature (IQ) signals, which are calibrated and transformed to obtain the Channel Impulse Response (CIR). Subsequently, the CIR data are processed using the Space Alternating Generalized Expectation Maximization (SAGE) algorithm to extract the channel parameters for each snapshot, including the propagation delay $\tau$, AOA $\theta$, and EOA $\phi$. The SAGE algorithm is a widely used high-resolution parameter estimation (HRPE) algorithm based on the expectation-maximization (EM) framework, with detailed descriptions found in \cite{fessler2002space}, \cite{chong2002joint}, and \cite{matthaiou2007characterization}. Finally, de-noising and spurious path filtering are applied to construct a set of channel feature vectors consisting of $N$ multi-path components (MPCs). This can be expressed as:
\begin{equation}
\Theta=\left\{\left(\tau_i, \theta_i, \phi_i\right)\right\}_{i=1}^N
\end{equation}

% The point cloud data acquired from LiDAR is first down-sampled to reduce environmental redundancy and alleviate training complexity. Principal Component Analysis (PCA) is then applied to normalize the 3D orientation of the point cloud data. Detailed descriptions of PCA can be found in \cite{hotelling1933analysis} and \cite{shlens2014tutorial}. Given that the LiDAR provides a 360-degree field of view while the ISAC channel has a limited sensing range, certain parts of the point cloud may contain information beyond the ISAC perception scope. Therefore, spatial cropping is essential to ensure spatial consistency between the point cloud and the channel data. Each frame of point cloud is denoted as:
The point cloud data acquired from LiDAR is first down-sampled to reduce environmental redundancy and alleviate training complexity. Subsequently, Principal Component Analysis (PCA) is applied to align the coordinate axes of the point cloud data. Since the raw LiDAR data is in a Cartesian ego-centric frame, PCA is utilized to align the principal axes—orienting the y-axis parallel to the vehicle's heading, the x-axis to the sensing direction, and the z-axis vertically. This correction addresses axis offsets caused by vehicle motion and ensures spatial consistency. Detailed descriptions of PCA can be found in \cite{hotelling1933analysis} and \cite{shlens2014tutorial}. Given that the LiDAR provides a 360-degree field of view while the ISAC channel has a limited sensing range, certain parts of the point cloud may contain information beyond the ISAC perception scope. Therefore, spatial cropping is essential to ensure spatial consistency between the point cloud and the channel data. Each frame of point cloud is denoted as:
\begin{equation}
\mathcal{P}=\left\{\mathbf{p}_j=\left(x_j, y_j, z_j\right)\right\}_{j=1}^M
\end{equation}
where $M$ represents the number of valid points in the point cloud and $\mathbf{p}_j$ denotes the 3D coordinates of the $j$-th sampled point.

Subsequently, the processed point cloud $\mathcal{P}$ and the set of channel parameters $\Theta$ are paired sample by sample based on timestamp and spatial location to ensure environmental consistency across the training data. Scene labels are manually annotated using panoramic RGB video frames in conjunction with the point cloud data, and are categorized into classes such as "scenes containing buildings or trees only", "scenes containing both buildings and trees". These labels support the multi-task training process in the proposed framework.

The model is trained on the constructed dataset to learn a mapping from channel characteristics to environmental representations:
\begin{equation}
\hat{\mathcal{P}}=f(\Theta)
\end{equation}
where $\Theta$ denotes the input set of channel characteristics, $\hat{\mathcal{P}}$ is the reconstructed 3D environment, and $f$ represents the proposed MSCR-Net, which will be elaborated in Section V.

In this work, we adopt point clouds, a widely used format in the robotics domain to represent the geometric structure of the reconstructed 3D environment \cite{lu2020see}. This representation has been utilized in recent learning-based 3D reconstruction studies \cite{qi2017pointnet,fan2017point,qi2017pointnet++}. A point cloud models an object as an unordered set of points sampled from its surface, each represented by its 3D Cartesian coordinates. Compared to voxel-based representations, point clouds offer high-resolution geometry encoding with significantly lower memory overhead.
%%%%%%%系统框图%%%%%%%%%%%%%
\begin{figure*}[t]
    \centering
    \includegraphics[width=\linewidth]{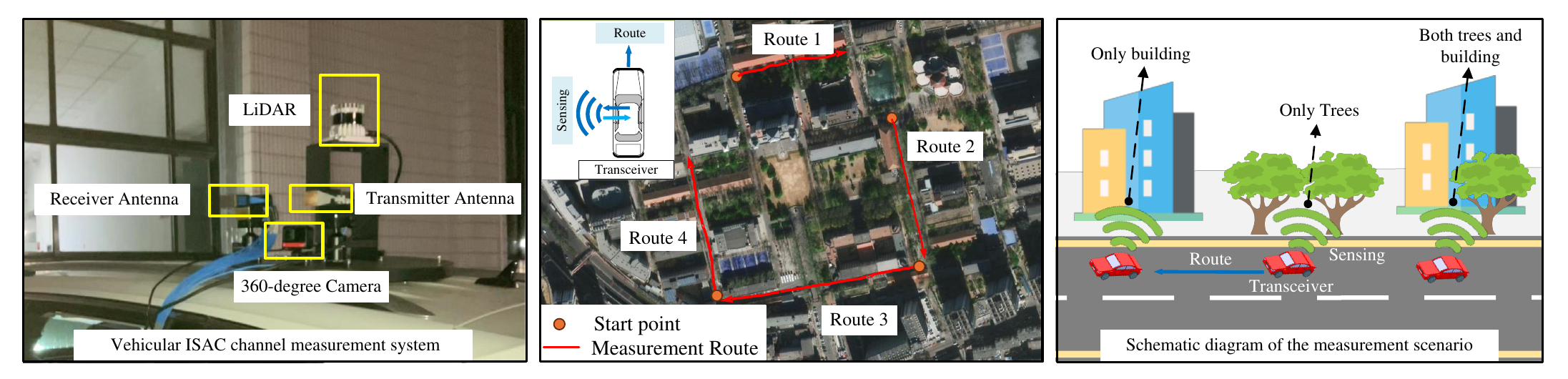}
    \caption{The measurement campaign.}
    \label{fig:I-R—S}
\end{figure*}
% \begin{figure*}[t]
%   \centering
%   \begin{minipage}[b]{0.32\textwidth}
%     \centering
%     \includegraphics[width=\textwidth]{设备-v2.pdf}
%     \caption*{(a)}
%   \end{minipage}
%   \hfill
%   \begin{minipage}[b]{0.32\textwidth}
%     \centering
%     \includegraphics[width=\textwidth]{路线-v2.pdf}
%     \caption*{(b)}
%   \end{minipage}
%   \hfill
%   \begin{minipage}[b]{0.32\textwidth}
%     \centering
%     \includegraphics[width=\textwidth]{场景-v2.pdf}
%     \caption*{(c)}
%   \end{minipage}
%   \caption{Illustration of three horizontally aligned figures.}
%   \label{fig:3figs}
% \end{figure*}

%%%%%%%%%%%%%%%%%%%%%%%%%%
% %%%%%%%设备%%%%%%%%%%%%%
% \begin{figure}[t]
%     \centering
%     \includegraphics[width=0.95\linewidth]{Installationv3.pdf}
%     \caption{Vehicular Measurement system}
%     \label{fig:Installation}
% \end{figure}

\section{ISAC CHANNEL MEASUREMENTS}%%%第三章
In this section, we present the vehicular ISAC channel measurement campaign conducted to validate the proposed reconstruction framework. We detail the construction of a high-precision multi-modal measurement platform capable of simultaneously capturing mmWave channel parameters and environmental ground truths. These real-world measurements serve as the fundamental basis for establishing the joint channel-environment dataset, ensuring that the subsequent deep learning model is trained on physically accurate propagation characteristics.
% %%%%%%%%%%%%%%%%%%%%%%%
\begin{table}[t]
\centering
\caption{Configurations of measurement system}
\begin{tabular}{lc}
\toprule
\textbf{Parameters} & \textbf{Value} \\
\midrule
Center frequency           & 28\,GHz \\
Bandwidth                  & 1\,GHz \\
Transmit power             & 28\,dBm \\
Sounding signal            & Multi-carrier signal \\
Number of frequency points & 1024 \\
Sample rate                & 6.4\,GHz \\
Transmitter antenna        & Horn antenna \\
Horizontal coverage angle (Tx)     & 20$^\circ$ \\
Vertical coverage angle (Tx)    & 20$^\circ$ \\
Receiver antenna           & 4$\times$8 array antenna \\

\bottomrule
\end{tabular}
\label{tab:config}
\end{table}
\subsection{Measurement Equipment}%%%测量设备
The vehicular channel measurement system for ISAC scenarios is designed with key components, as shown in Fig. \ref{fig:I-R—S}. These include a signal generator-based TX, a signal digitizer-based RX, a LiDAR, and a 360-degree camera. This setup enables simultaneous sensing of the electromagnetic and physical environments for comprehensive situational awareness. The TX employs a signal generator, the National Instruments (NI) PXIe-5745 (also known as the PXI FlexRIO Signal Generator). The RX employs a signal analyzer, the NI PXIe-5775 (also known as the PXI FlexRIO Digitizer). The probing signal is a multi-carrier waveform with a center frequency of 28 GHz and a bandwidth of 1 GHz. For antenna configurations, the Tx is equipped with a horn antenna, while the Rx uses a 4×8 antenna array. To ensure synchronization between the Tx and Rx, both ends are disciplined by GPS-referenced rubidium clocks providing a 10 MHz reference pulse. The detailed configuration of the measurement system is summarized in TABLE \ref{tab:config}. For physical dynamic environment sensing, a LiDAR and 360-degree panoramic camera are positioned with the TX/RX, capturing point cloud data, RGB video, and channel data simultaneously.
\begin{figure}[t]
\centering
\subfigure[]{\includegraphics[width=1.7in]{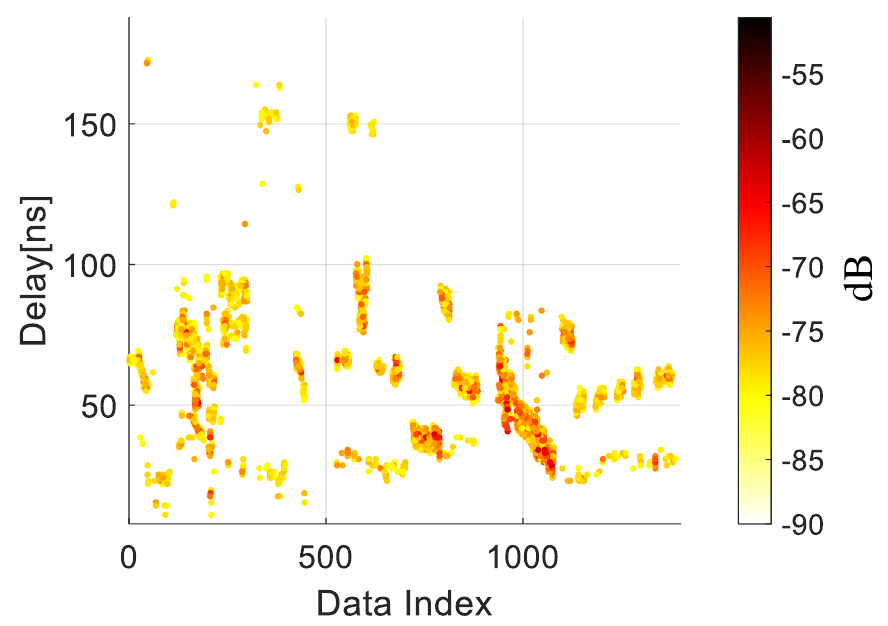}}
\subfigure[]{\includegraphics[width=1.7in]{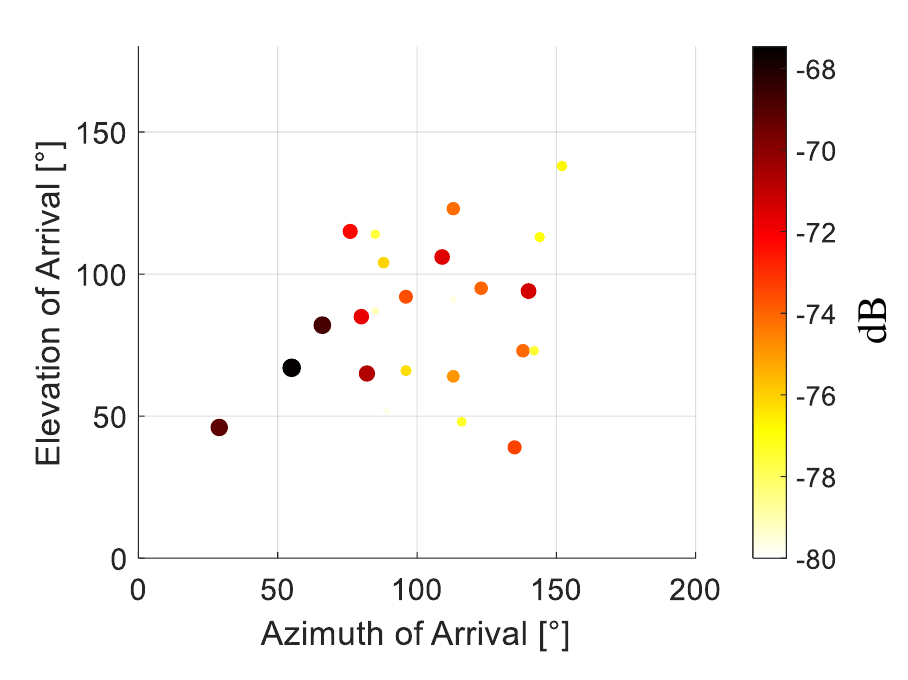}}
\caption{Channel parameters extracted by SAGE. (a) Delay. (b) AOA and EOA.}
\label{fig:SAGE}
\end{figure}

It is important to note that while this specific measurement setup is specialized, the employed SAGE algorithm effectively de-embeds the antenna patterns during parameter extraction. Consequently, the resulting channel parameters represent the intrinsic propagation environment, independent of the specific measurement hardware. The primary objective of this setup is channel sounding rather than practical beamforming for data transmission. In channel sounding, using a Tx horn antenna ensures that the probing signal illuminates a broader spatial area to capture all potential environmental multipath components comprehensively, whereas practical directional beamforming would spatially filter out many reflectors. The main advantage of this setup, combined with the planar Rx array, is its ability to extract highly stable and accurate angular and delay resolutions of the intrinsic environment without being biased by a narrow transmit beam. However, the limitation is that this sounding-focused setup does not reflect the hardware imperfections or dynamic beam-sweeping overhead inherent in practical commercial ISAC transceivers.
%%%%%%%测量系统%%%%%%%%%%%%%
% \begin{figure}[htbp]
%   \centering
%   \begin{minipage}[b]{0.35\linewidth}
%     \centering
%     \includegraphics[width=\linewidth]{Installationv2.pdf}
%     \captionof{figure}{Measurement system.}
%     \label{fig:Installation}
%   \end{minipage}
%   \hfill
%   \begin{minipage}[b]{0.61\linewidth}
%     \centering
%     \begin{tabular}{lc}
%       \toprule
%       Parameters & Value \\
%       \midrule
%       Center frequency     & 28\,GHz \\
%       Bandwidth            & 1\,GHz \\
%       Transmit power       & 28\,dBm \\
%       Sounding signal      & Multi-carrier \\
%       Transmitter antenna  & Horn antenna \\
%       Sensing antenna      & 4$\times$8 array \\
%       Transceiver height   & 2.3\,m \\
%       Sample rate          &6.4\,GHz \\
%       \bottomrule
%     \end{tabular}
%     \captionof{table}{Configurations of measurement system}
%     \label{tab:config}
%   \end{minipage}
%   \vspace{0em}
%   \caption*{\textbf{Fig.\ref{fig:Installation} \& TABLE~\ref{tab:config}.} Measurement system.}
% \end{figure}

% %%%%%%%测量场景%%%%%%%%%%%%%
% \begin{figure}[t]
%     \centering
%     \includegraphics[width=0.95\linewidth]{scenev2.pdf}
%     \caption{Measurement Routes}
%     \label{fig:scene}
% \end{figure}
% %%%%%%%%%%%%%%%%%%%%%%%%%%
%%%%%%%%%%%%%%%%%%%%%%%%%%
\begin{figure}[t]
	\centering
	\subfigure[]{\includegraphics[width=1.1in]{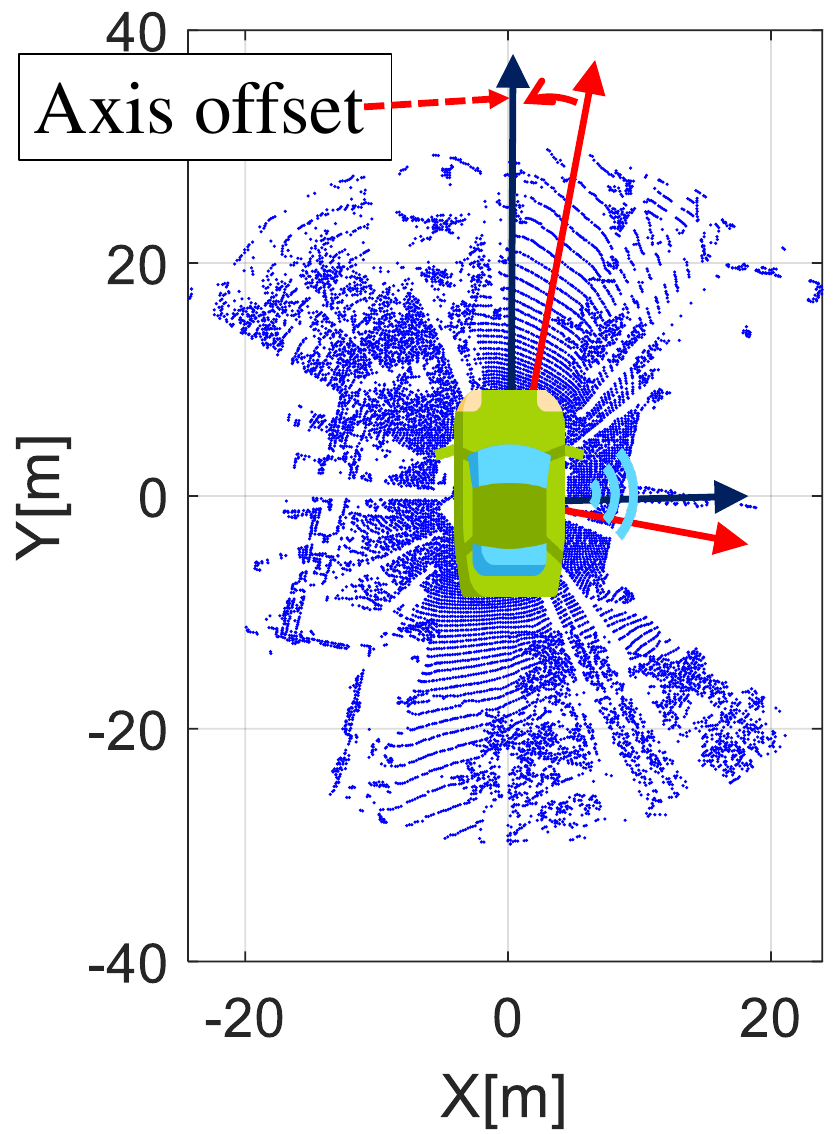}}
	\hfill
	\subfigure[]{\includegraphics[width=2.3in]{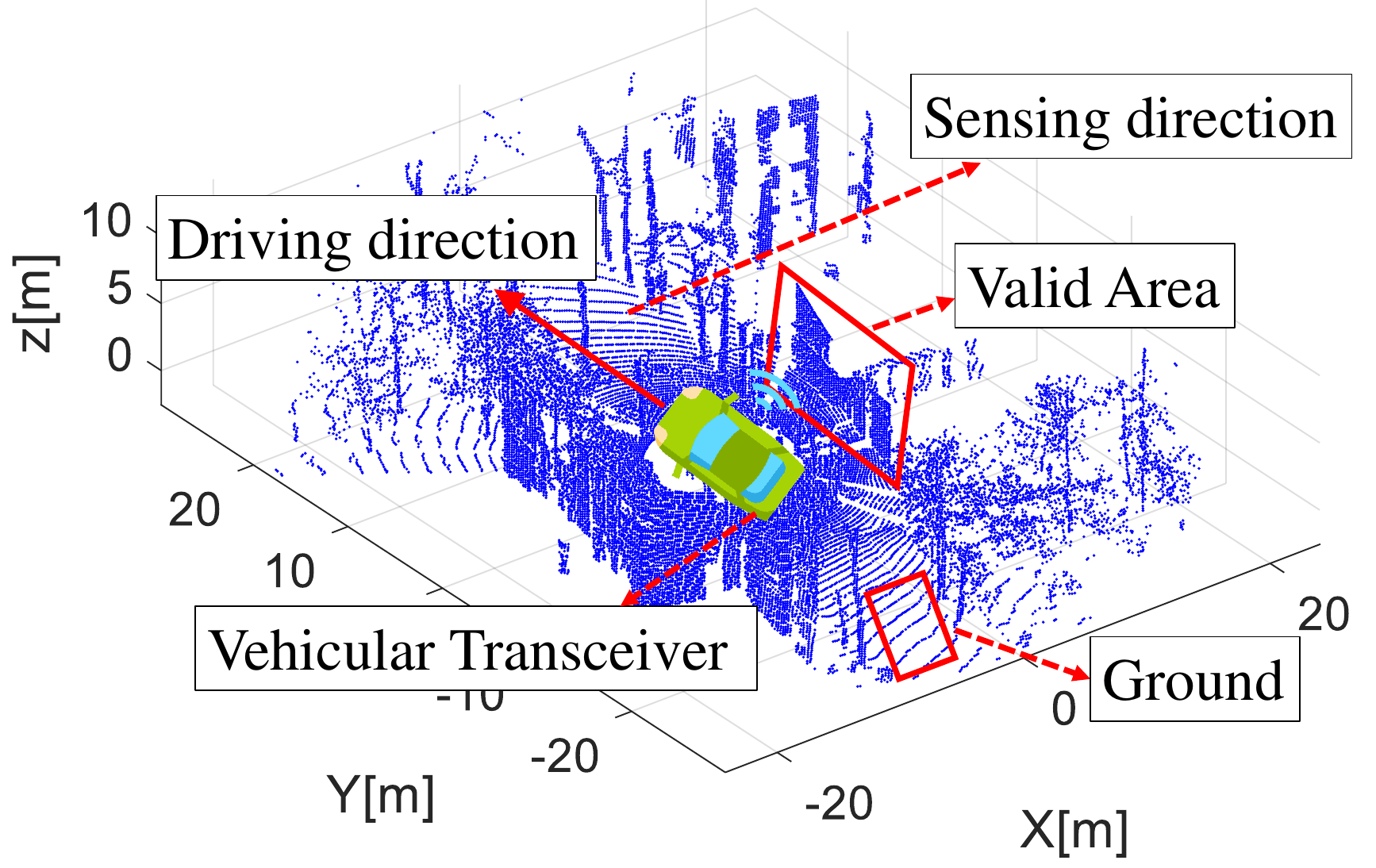}}
	\\
	\subfigure[]{\includegraphics[width=1.1in]{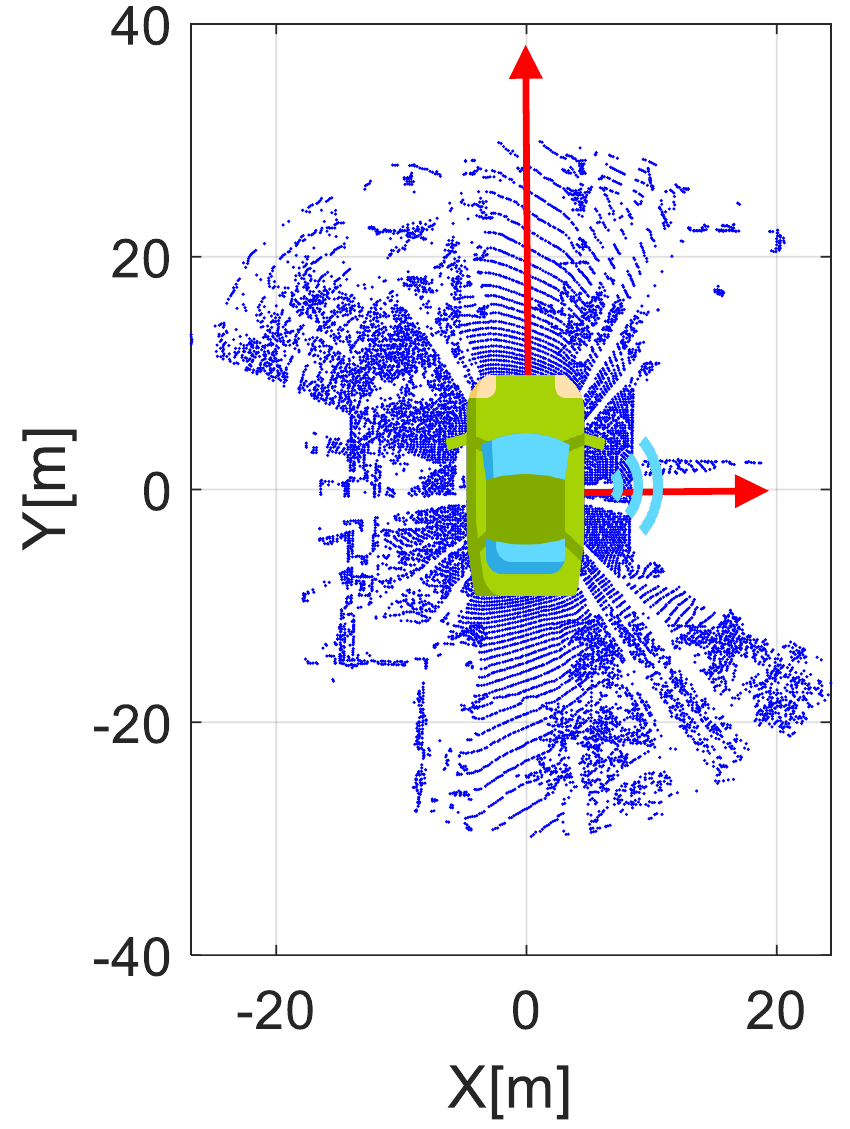}}
	\hfill
	\subfigure[]{\includegraphics[width=2.3in]{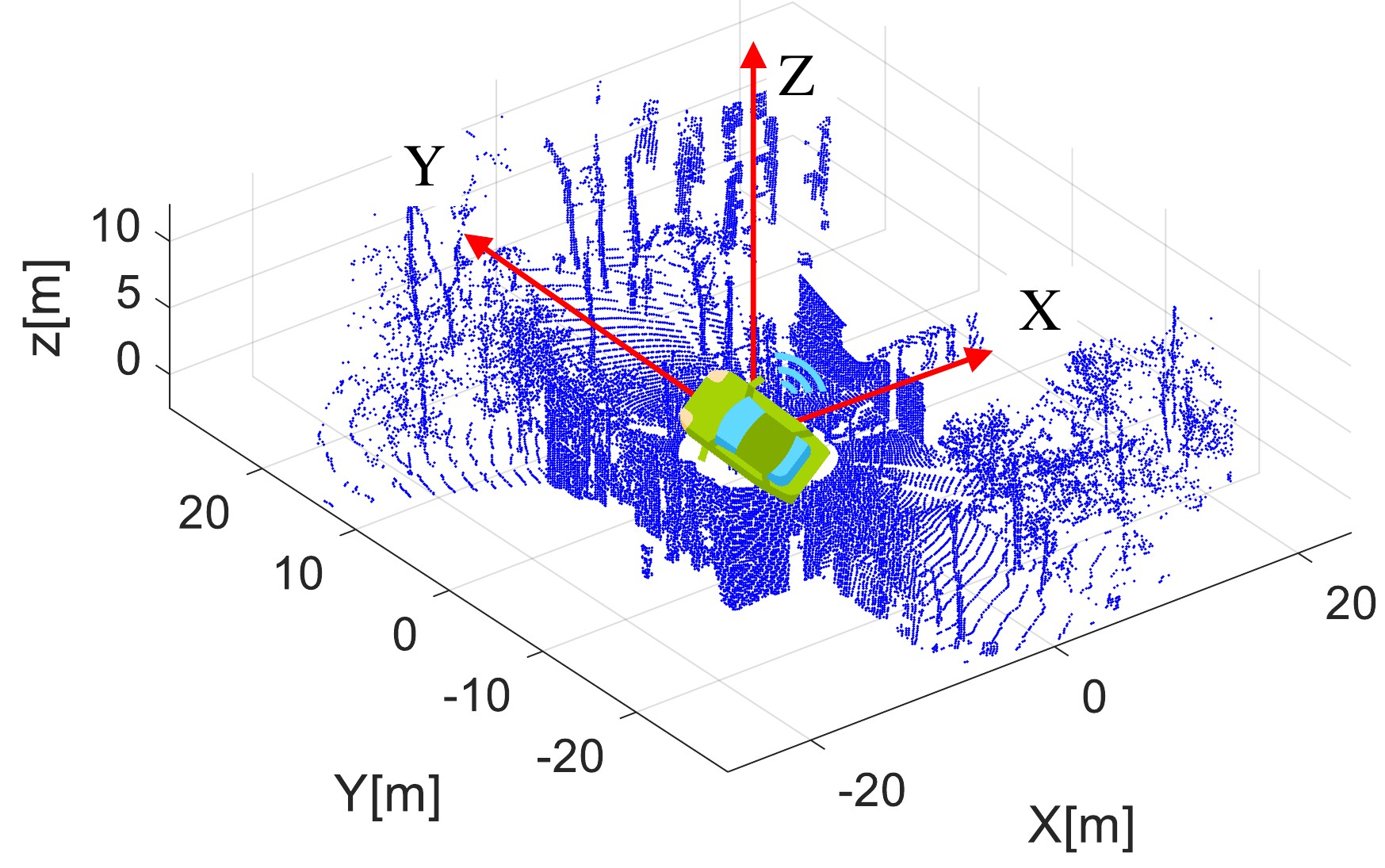}}
        \\
        \subfigure[]{\includegraphics[width=1.1in]{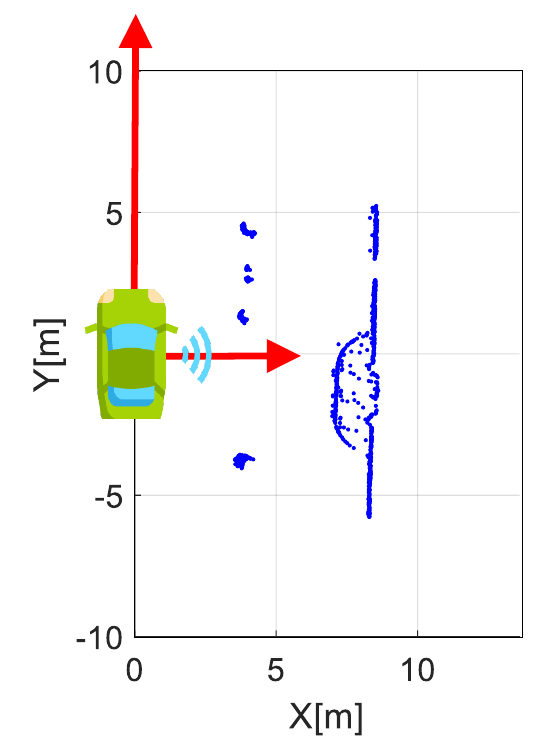}}
	\hfill
	\subfigure[]{\includegraphics[width=2.34in]{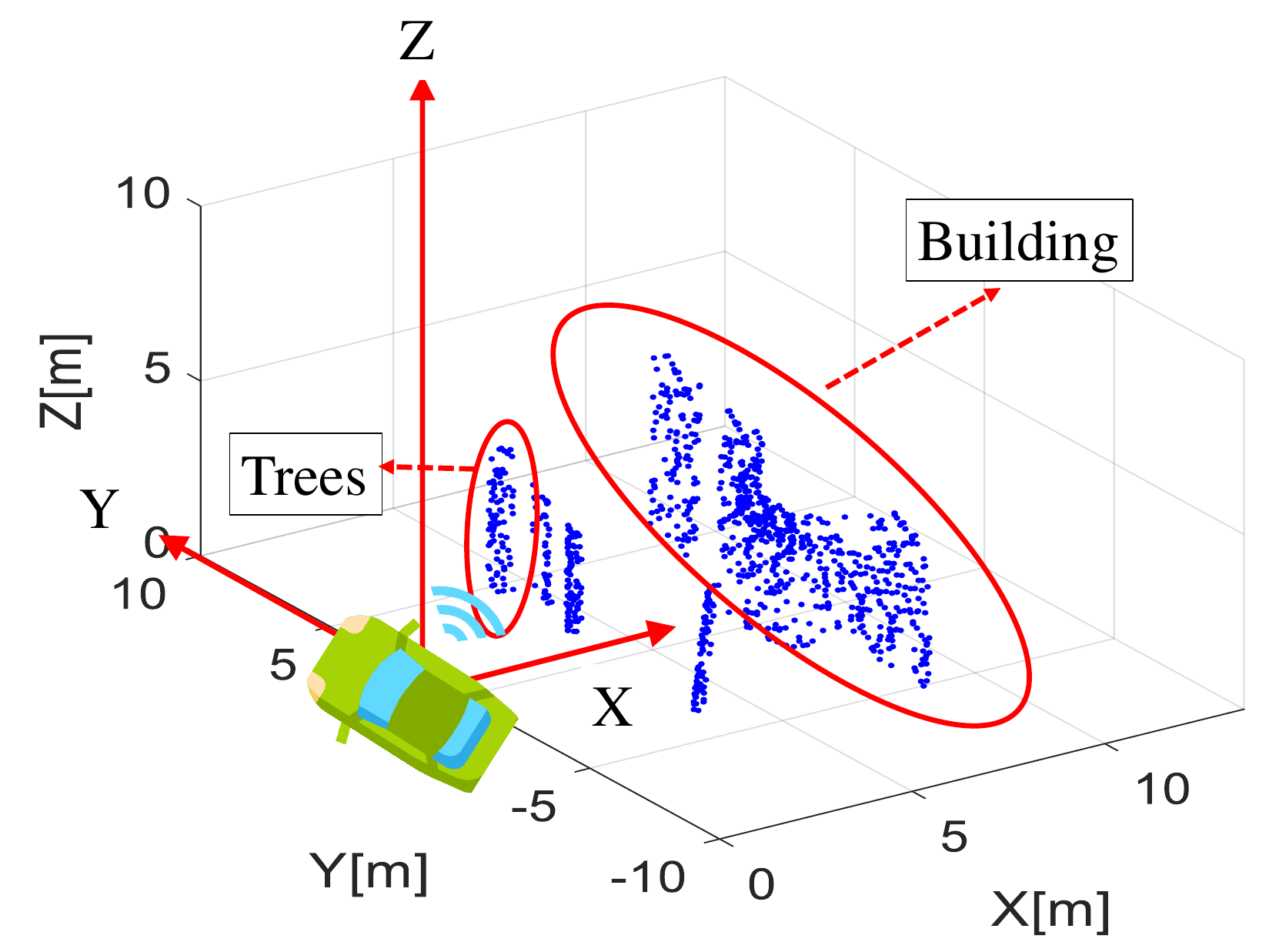}}
	\caption{LiDAR point cloud pre-processing. (a) The xy-plane of the point cloud of a frame. (b) One frame of point cloud. (c) The xy-plane of the point cloud of a frame after PCA. (d) One frame of point cloud after PCA. (e) The xy-plane of the point cloud of a frame after cropping. (f) One frame of point cloud after cropping.}
	\label{fig:datapreprocess}
\end{figure}
\subsection{Measurement Campaign}%%%测量场景
To construct ISAC channel characteristics and the corresponding LiDAR point cloud dataset, we conducted a field measurement campaign at Beijing Jiaotong University in Beijing, China. The measurement campaign covers four test routes, as illustrated in Fig. \ref{fig:I-R—S}. The test vehicle moved at a constant speed of approximately 10 km/h in a clockwise direction. The sensing direction of the channel measurement equipment is fixed to the left side of the vehicle's motion. This side-looking configuration was specifically selected to optimize the 3D reconstruction of roadside infrastructure (e.g., street canyons and building facades) for high-precision mapping tasks. Unlike front-facing setups where angular resolution degrades towards the vanishing point, the side-facing orientation allows the system to scan the environmental profile orthogonal to the vehicle's motion vector, capturing a continuous and high-resolution layout of parallel static structures. 

Consequently, the measurement focuses primarily on the outer perimeter of the sensing loop, capturing a variety of typical urban road scenarios, as illustrated in Fig. \ref{fig:I-R—S}. These scenarios consider various spatial configurations of buildings and trees (e.g., buildings only, trees only, and coexistence of buildings and trees), simulating the sensing of roadside environments as a vehicle travels through the scene. Specifically, Route 2 features a scene with only roadside trees, characterized by irregular distribution and non-deterministic channel reflections. Route 3 includes a scene bordered only by buildings, where the channel exhibited strong, continuous, and structured reflections. In addition, all routes contain composite scenes with both trees and buildings. Compared to buildings, trees were generally positioned closer to the vehicle, resulting in more diverse sensing paths and complex structural channel characteristics. This combination of heterogeneous environments effectively simulates common perception tasks in V2X and autonomous driving systems, providing a solid foundation for evaluating the model's generalization capability across different environmental conditions.
% The proposed method is designed primarily to meet the practical application needs of vehicular ISAC systems in dynamic urban traffic environments, encompassing typical scenarios such as V2X sensing, autonomous driving, and object recognition. A representative urban road simulation scenario is illustrated in Fig. \ref{fig:layout}. This scenario considers various spatial configurations of buildings and trees (e.g., buildings only, trees only, and coexistence of buildings and trees), simulating the sensing of roadside environments as a vehicle travels through the scene.
%%%%%%%相机%%%%%%%%%%%%%
\begin{figure}[t]
	\centering
	\subfigure[]{\includegraphics[width=1.7in]{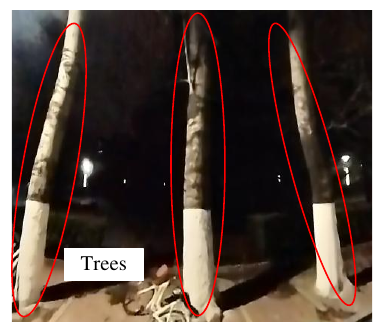}}
	\hfill
	\subfigure[]{\includegraphics[width=1.7in]{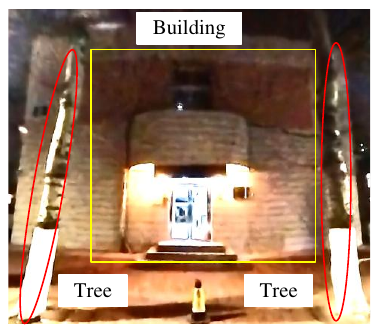}}
        \\
        \subfigure[]{\includegraphics[width=1.7in]{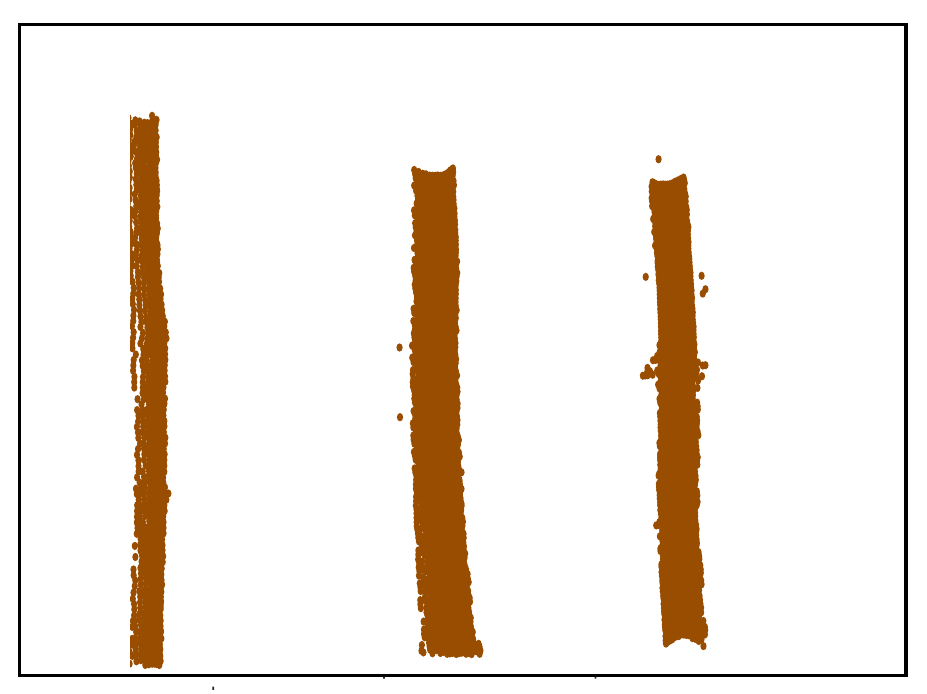}}
	\hfill
	\subfigure[]{\includegraphics[width=1.7in]{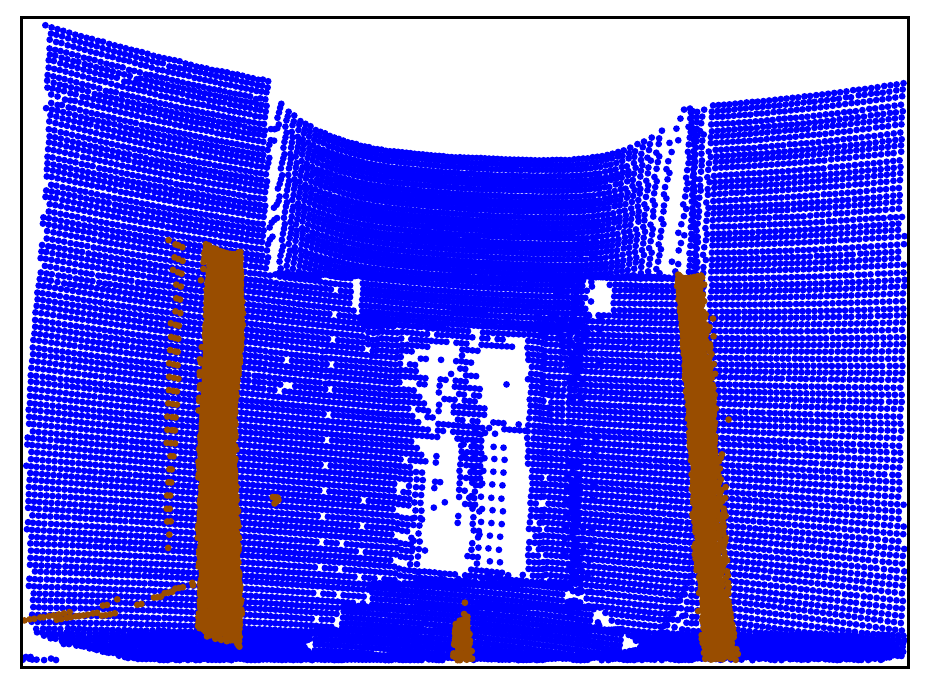}}
	\caption{RGB and LiDAR data of different scene categories. (a) Scene containing trees only. (b) Scene containing both buildings and trees. (c) LiDAR point cloud corresponding to Fig. \ref{fig:camera}(a). (d) LiDAR point cloud corresponding to Fig. \ref{fig:camera}(b).}
	\label{fig:camera}
\end{figure}
%%%%%%%%%%%%%%%%%%%%%%%%%%
\section{DATA PRE-PROCESSING}
In this section, we detail the comprehensive data pre-processing pipeline designed to transform the raw multi-modal measurements into a high-quality, unified dataset suitable for deep learning training. As the raw signals captured by different sensors vary significantly in format, sampling rate, and coordinate systems, a systematic processing workflow is essential. We first elaborate on the extraction of intrinsic channel parameters using the SAGE algorithm, followed by the rigorous spatiotemporal alignment and cropping of LiDAR point clouds. Finally, we describe the semantic labeling strategy and the partitioning of the dataset, which ensures the reliability and generalization capability of the subsequent environment reconstruction model.

\subsection{Multi-path Estimation}
The Channel parameters are extracted from the measured raw data by first processing the IQ signals to obtain the CIR. Subsequently, the SAGE algorithm, configured with Serial Interference Cancellation (SIC) and a model order of $L=50$, is employed to jointly estimate the MPCs (delay, AoA, and EoA) on a single-snapshot basis to preserve dynamic environmental features. The estimation utilizes a rectangular window and terminates after 5 iterations or upon convergence, after which the top-40 MPCs ranked by power magnitude are selected as the final input features. To enhance data integrity and improve the reliability of environment reconstruction, outliers are identified and removed. Fig. \ref{fig:SAGE} illustrates the distribution of pre-processed multi-path channel parameters across different dimensions. Specifically, Fig. \ref{fig:SAGE}(a) shows the distribution of path delays across multiple snapshots, while Fig. \ref{fig:SAGE}(b) presents the angular distribution of path directions of arrival (DoA) for a single snapshot. Each scatter point in the figures represents a valid multi-path component estimated at the corresponding snapshot, with the color indicating the power attenuation experienced by the signal during propagation through the channel.
%%%%%%%%%%%%%%%%%%%%%%%%%%
%%%%%%%模型%%%%%%%%%%%%%
\begin{figure*}[t]
    \centering
    \includegraphics[width=\linewidth]{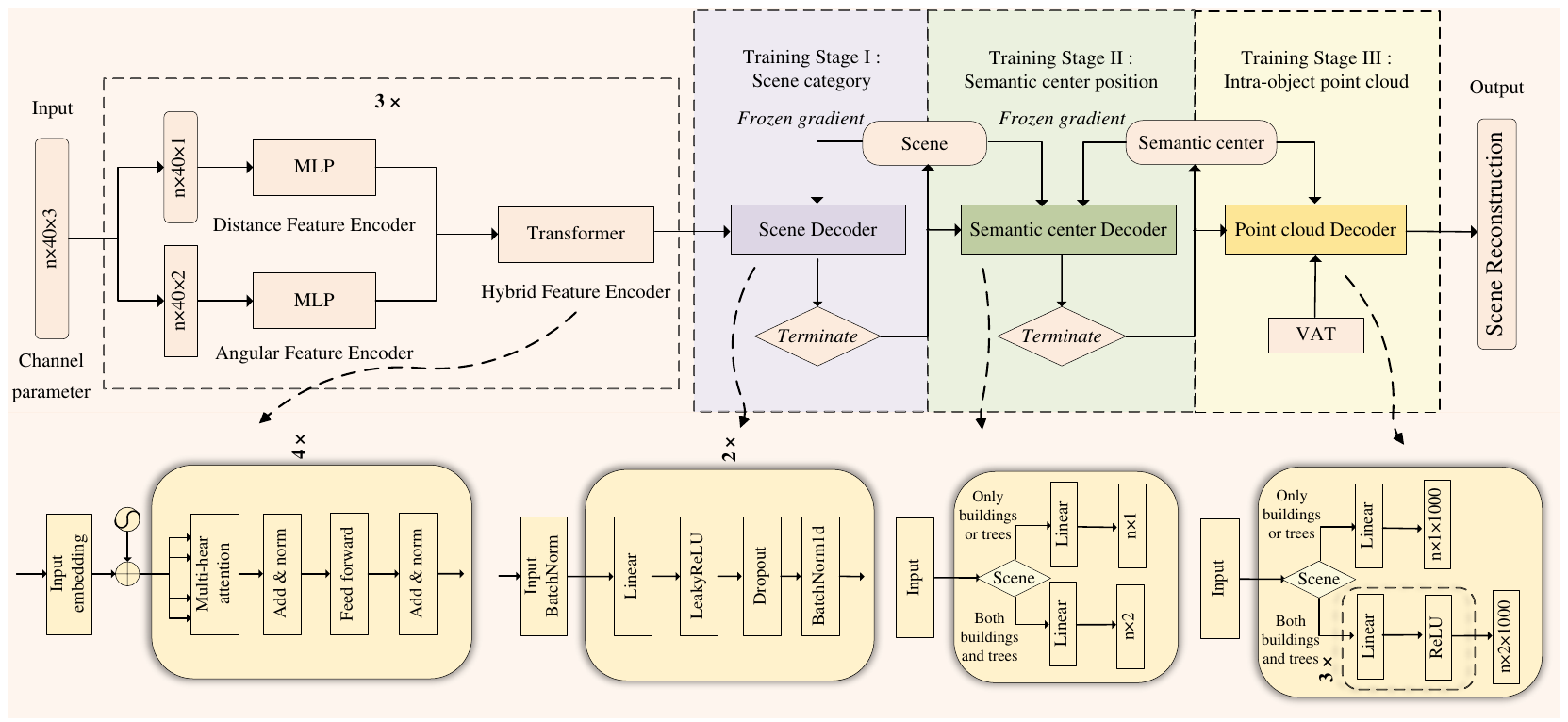}
    \caption{The proposed MSCR-Net.}
    \label{fig:model}
\end{figure*}
%%%%%%%%%%%%%%%%%%%%%%%%%%
\subsection{LiDAR Data Pre-processing}
LiDAR point cloud data are subjected to three main steps: synchronization, alignment, and cropping. First, temporal synchronization is achieved by uniformly downsampling the 10 Hz LiDAR stream to match the 2 Hz channel acquisition rate. Subsequently, PCA is employed to align the coordinate system—orienting the y-axis to the vehicle's heading and the x-axis to the sensing direction—followed by manual fine-tuning to correct residual rotational offsets. Finally, to ensure Field-of-View (FOV) consistency with the ISAC system, the point cloud is spatially cropped to match the Tx horn antenna's beamwidth, specifically limiting the valid volume to within $\pm 10^{\circ}$ in both azimuth and elevation. Finally, the exact 1000-point sampling procedure involves uniformly downsampling the variable-length cropped point cloud for each snapshot to a fixed size of 1000 points to facilitate batch training. As shown in Fig. \ref{fig:datapreprocess}(b), the point cloud represents the surrounding environment of the vehicular transceiver. Due to the non-omnidirectional sensing range of the ISAC channel, environmental elements outside the valid sensing area, such as ground points and point clouds from the non-sensing direction, are considered unnecessary. In addition, as illustrated in Fig. \ref{fig:datapreprocess}(a), the axis offset caused by the vehicle's heading deviation must be corrected via PCA to ensure spatial alignment. The processed point cloud data after these steps is shown in Fig. \ref{fig:datapreprocess}(f).
%%%%%%%SAGE%%%%%%%%%%%%%
% \begin{figure}[t]
% 	\centering
% 	\subfloat[]{\includegraphics[width=1.74in]{SAGE2.pdf}%
% 		\label{fig:delay}}
% 	\hfill
% 	\subfloat[]{\includegraphics[width=1.74in]{SAGE2v5.pdf}%
% 		\label{fig:AOAEOA}}
% 	\caption{Channel parameters extracted by SAGE. (a) Delay.  (b) AOA and EOA.}
% 	\label{fig:SAGE}
% \end{figure}

%%%%%%%LIDAR%%%%%%%%%%%%%
% \begin{figure}[!t]
% 	\centering
% 	\subfloat[]{\includegraphics[width=1.14in]{datapreprocessv2.pdf}%
% 		\label{Fig.2.1GHz}}
% 	\hfill
% 	\subfloat[]{\includegraphics[width=2.34in]{datapreprocess-bv3.pdf}%
% 		\label{Fig.2.6GHz}}
% 	\\
% 	\subfloat[]{\includegraphics[width=1.14in]{datapreprocess-cv2.pdf}%
% 		\label{Fig.3.5GHz}}
% 	\hfill
% 	\subfloat[]{\includegraphics[width=2.34in]{datapreprocess-dv2.pdf}%
% 		\label{Fig.omni}}
%         \\
%         \subfloat[]{\includegraphics[width=1.14in]{datapreprocess-ev2.pdf}%
% 		\label{Fig.2.1GHz}}
% 	\hfill
% 	\subfloat[]{\includegraphics[width=2.34in]{datapreprocess-fv2.pdf}%
% 		\label{Fig.2.6GHz}}
% 	\caption{LiDAR point cloud pre-processing. (a) The xy-plane of the point cloud of a frame.  (b) One frame of point cloud. (c) The xy-plane of the point cloud of a frame after PCA.  (d) One frame of point cloud after PCA.  (e) The xy-plane of the point cloud of a frame after cropping.  (f) One frame of point cloud after cropping.}
% 	\label{fig:datapreprocess}
% \end{figure}

\subsection{Dataset Labeling}
After pre-processing, channel data and LiDAR point clouds are aligned frame by frame and labeled according to the types of scatterers present in the environment, which is essential for the staged training of the proposed model in Section V. The manual annotation was performed by two independent personnel using MATLAB as the labeling tool. Specifically, the labeling rule is based on observing the presence of trees and buildings in the pre-processed LiDAR data, classifying the scene of each sample into two categories: (1) scenes containing buildings or trees only, and (2) scenes containing both buildings and trees. Because the pre-processed LiDAR data provides sufficient resolution for the surveyed urban scenarios, there is practically no ambiguity in distinguishing between buildings and trees, making the labeling process largely unaffected by subjective human bias. For the consistency check method, since the LiDAR and channel data are strictly aligned frame-by-frame, the process primarily involved verifying the categorical accuracy of the labels assigned by the annotators. The scene categories for each sample are also cross-referenced using RGB video data acquired from a panoramic camera. Fig. \ref{fig:camera} illustrates the real-world scenes corresponding to two representative samples. Specifically, Fig. \ref{fig:camera}(a) shows a sample categorized as a scene containing trees only, while Fig. \ref{fig:camera}(b) depicts a scene containing both buildings and trees. Based on this classification, we further define a "semantic object" as a single building facade or a group of trees. In vehicular dynamic scenarios, trees are typically located in front of buildings. It is important to note that the building facades almost always face the direction of channel sensing; thus, the term "building" in this paper specifically refers to the facade of a building. These labels effectively capture the structural diversity and complexity of the environment, providing the model with more discriminative training samples.
%%%%%%%%%%消融结果%%%%%%%%%%
\begin{figure*}[!t]
    \centering
    \includegraphics[width=\linewidth]{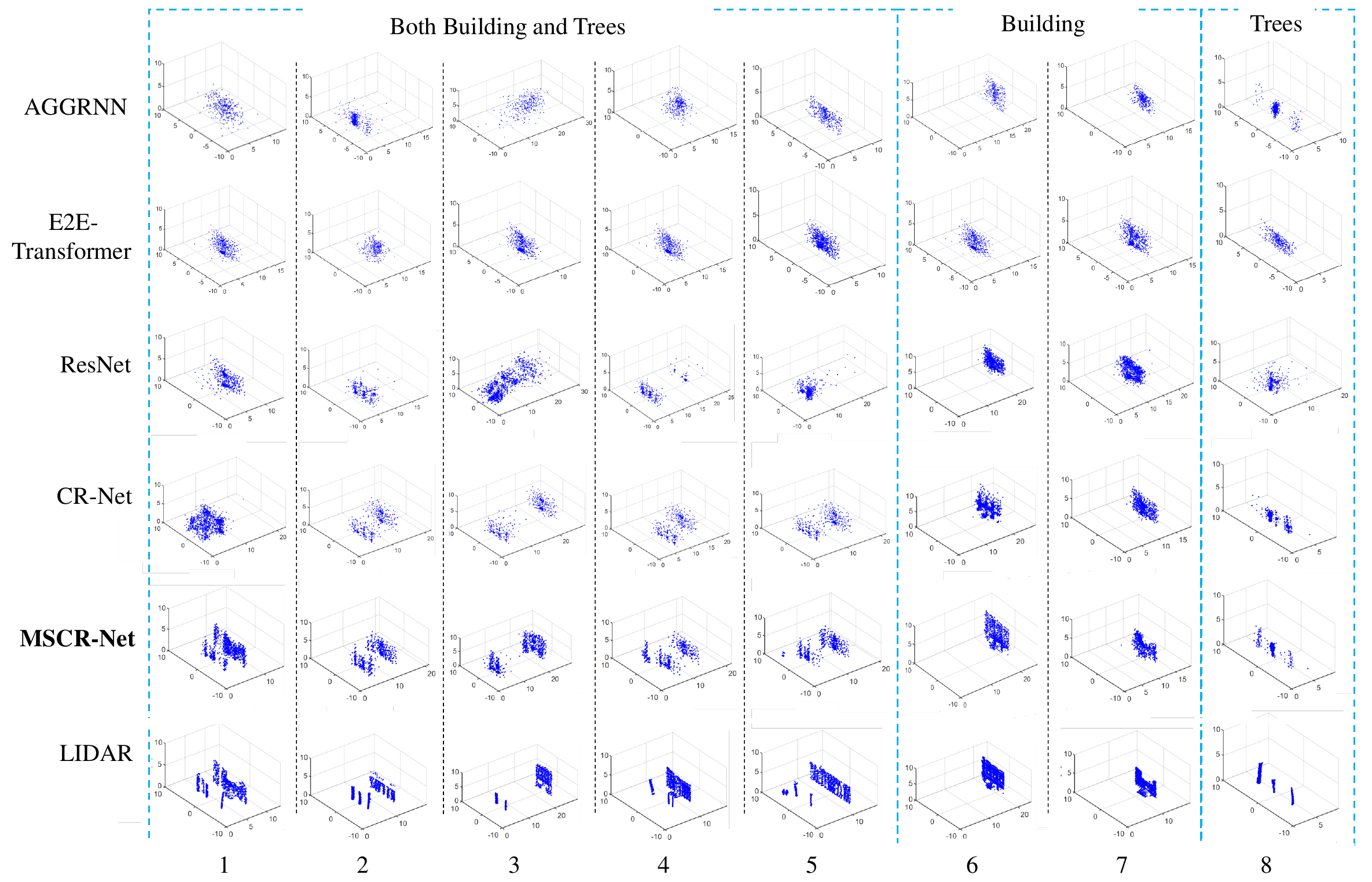}
    \caption{Comparison against different methods.}
    \label{fig:result-singlev2}
\end{figure*}
%%%%%%%%%%%%%%%%%%%%%%%%%%
\subsection{Dataset Partitioning}
Each pre-processed channel data sample has a size of 3x40, representing 40 multi-path components characterized by their delay, AoA, and EoA. The corresponding point cloud data have a size of 1000x3, representing 1000 3D coordinates within the environment. To ensure generalization across different environmental settings, the dataset maintains a roughly balanced number of samples across the two defined scene categories. A total of 4000 sample pairs are collected. Among them, data from Routes 1 through 3 are used to construct the training and validation sets. Specifically, 90\% of the samples from these routes are used for training, while the remaining 10\% are randomly selected as the validation set for model selection. Route 4, which represents a distinct path not seen during the training phase, is reserved as the test set to evaluate the final generalization performance.
\section{MODEL DESIGN}

Due to the differences in scattering characteristics between ISAC mmWave signals and LiDAR signals, their environmental sensing capabilities vary significantly. As a result, directly constructing an end-to-end network that maps ISAC channels to point clouds proves to be highly challenging in practice. However, by leveraging the inherent spatial consistency between channel parameters and surrounding environments, it becomes feasible to progressively decouple the spatial structure of the environment from the sparse channel characteristics. This process is analogous to the gradual perception of a scene by a human visual system: first identifying the number of objects, then locating their centers, and finally refining their shapes.

Motivated by this observation, a multi-stage training and reconstruction model based on MSCR-Net is proposed. The model consists of three structurally identical but independently parameterized feature encoders and a three-stage serial decoder. Together, they form three consecutive training phases, responsible for inferring scene category, semantic center position, and point cloud details, respectively. The general architecture of the model is illustrated in Fig. \ref{fig:model}. In the following sections, we elaborate on model design, workflow, loss function formulation, and training strategy.
%%%%%%%%%%%%%%CD-1%%%%%%%%%%%%
\begin{figure}[t]
    \centering
    \includegraphics[width=0.95\linewidth]{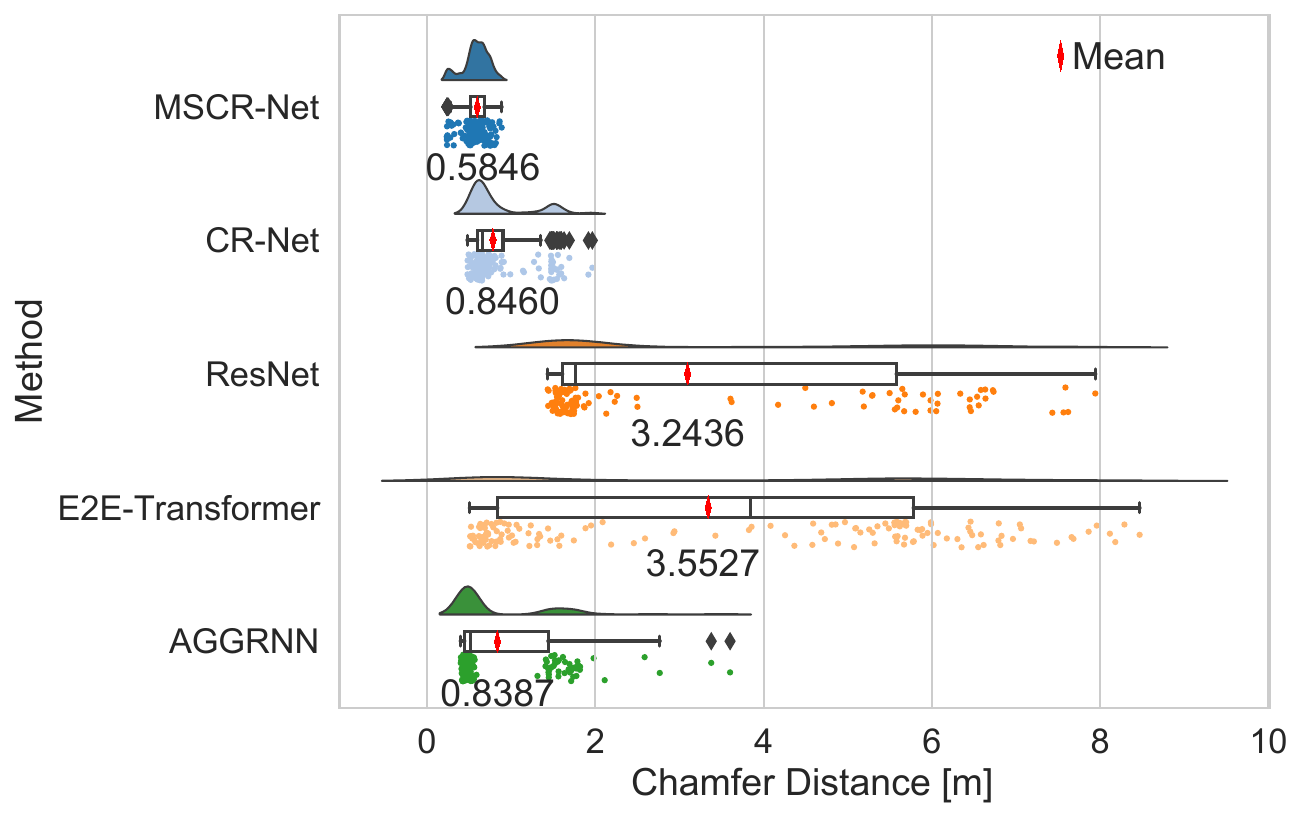}
    \caption{Chamfer Distance comparison against different methods.}
    \label{fig:CD-1}
\end{figure}
%%%%%%%%%%%%%%%%%%%%%%%%%%%%%%
\subsection{Model Structure}
The feature encoder comprises three components: a distance feature encoder, an angle feature encoder, and a hybrid feature encoder. The distance and angle encoders are each implemented as two independent multi-layer perceptrons (MLPs), responsible for processing delay and angular characteristics, respectively. The hybrid feature encoder consists of an input feature fusion module, a four-layer Transformer architecture, a global pooling layer, and a linear projection layer. Each Transformer layer includes a multi-head self-attention mechanism with four attention heads, a feedforward neural network, residual connections, and layer normalization. This structure facilitates the deep integration and encoding of delay and angular characteristics, ultimately generating a high-dimensional fused channel representation.

The feature decoder consists of three modules: a scene decoder, a semantic center decoder, and a point cloud decoder.

\noindent\textbf{1) Scene Decoder:} The scene decoder adopts a multi-layer fully connected neural network structure, including an input normalization layer followed by two fully connected layers. It predicts the scene category from the fused channel characteristics. The predicted category is represented in binary form (1 or 2), where 1 indicates scenes containing either buildings or trees, and 2 indicates scenes containing both buildings and trees.

\noindent\textbf{2) Semantic center Decoder:} The semantic center decoder is a fully connected network that selects appropriate linear layers based on the output of the scene decoder, and predicts the 3D coordinates of semantic object centers within the environment.

\noindent\textbf{3) Point cloud Decoder:} The point cloud decoder uses the semantic center predictions to generate point cloud distributions, outputting a total of 1000 3D points to represent the reconstructed environment. Given that mixed scenes containing both buildings and trees pose greater challenges for point cloud generation, we specifically design a dedicated sub-network composed of three MLP layers to improve reconstruction performance in such cases.

\subsection{Workflow}

The input of three feature encoders consists of multi-dimensional channel parameters, where each sample contains MPCs characterized by delay, AOA, and EOA. The fused channel characteristics generated by the encoders are fed into the decoders of the three prediction stages. The design objective is to freeze the gradient updates of the encoder corresponding to each current stage, ensuring that the training of subsequent stages remains unaffected.

In addition, the decoders in the semantic center prediction stage and point cloud prediction stage require auxiliary input from the output of the preceding stages. Specifically, the output of the scene decoder guides the semantic center decoder in generating an appropriate number of semantic object centers. The output of the semantic center decoder provides the spatial coordinates of semantic object centers, which are then used by the point cloud decoder to generate localized point cloud distributions.

It is worth noting that when information is passed from the first stage to the second, the output format must be transformed. Since binary outputs hinder effective gradient propagation, we adopt a probabilistic output form for scene category classification. The final binary label (1 or 2) is determined only after completing the inference in the first stage and is then passed to the semantic center decoder for further processing.
%%%%%%%%%%%%%%CD-2%%%%%%%%%%%%
\begin{figure}[t]
    \centering
    \includegraphics[width=0.95\linewidth]{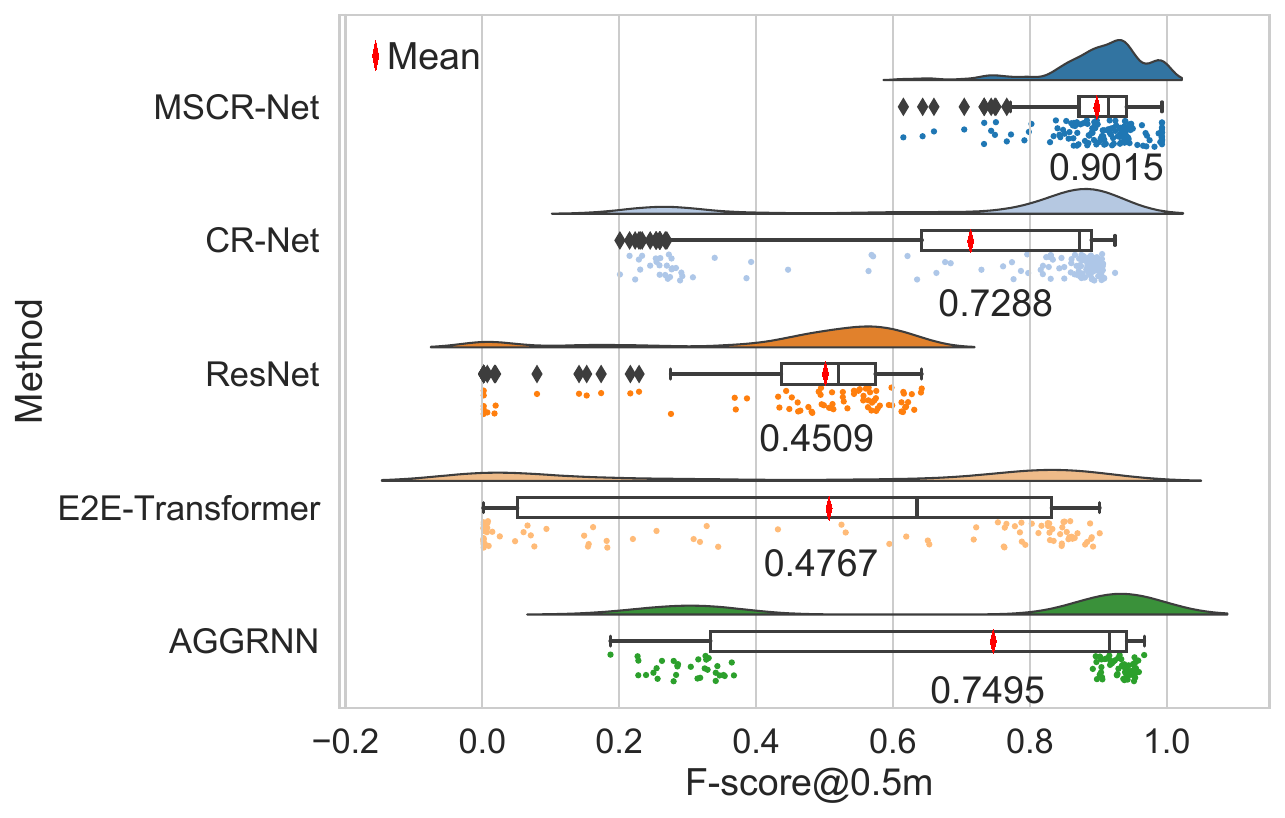}
    \caption{F-score@0.5m comparison against different methods.}
    \label{fig:CD-2}
\end{figure}
%%%%%%%%%%%%%%CD-2%%%%%%%%%%%%
\subsection{Loss Functions and Training Strategies}
In the multi-stage serial prediction architecture, the output of each preceding stage plays a critical role in training the subsequent stage. Accurate intermediate predictions serve as informative inputs, enabling the next-stage network to more effectively identify gradient descent directions and accelerate convergence. To support this process, we design dedicated loss functions and training strategies for each of the three prediction stages.
\begin{figure}[t]
\centering
\subfigure[]{\includegraphics[width=3.3in]{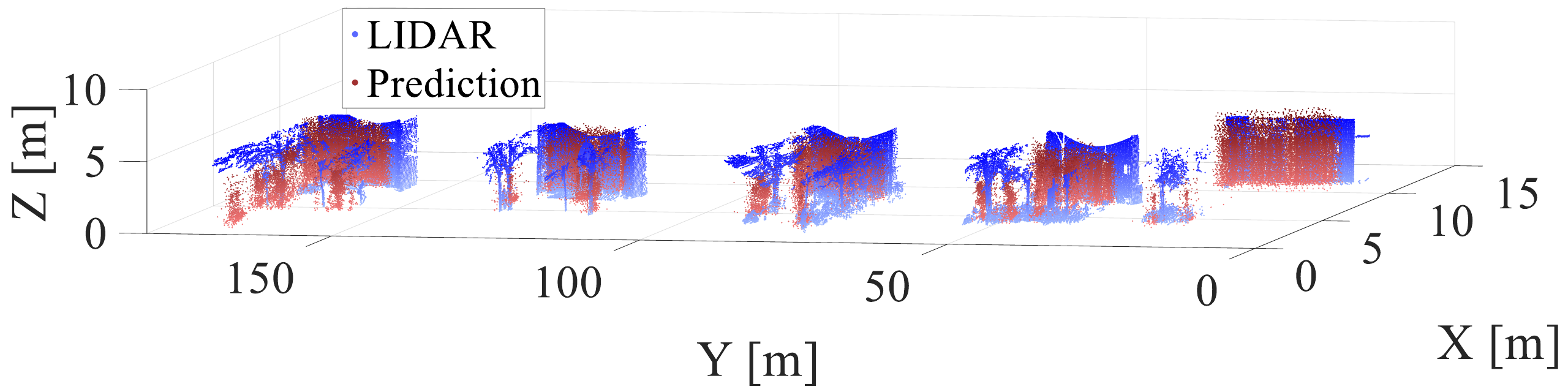}}
\subfigure[]{\includegraphics[width=3.3in]{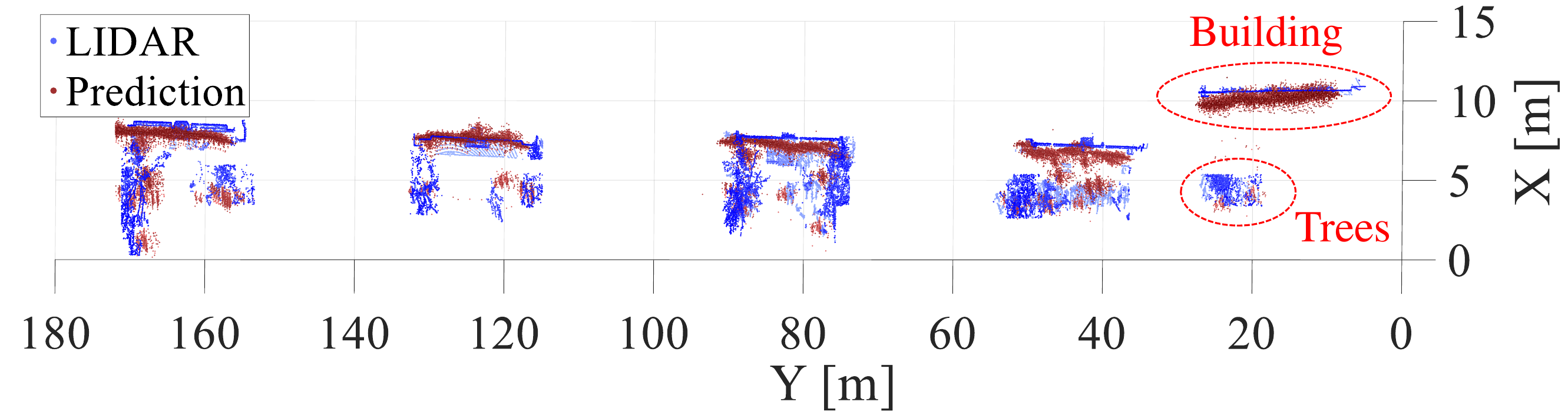}}
\caption{Comparison of proposed method with LiDAR. (a) Comparison of reconstructed scene and LiDAR ground truth in a unified coordinate system. (b) Comparison of proposed method with LiDAR.}
\label{fig:12}
\end{figure}
\subsubsection{Training Stage I: Scene category}
We adopt a cross-entropy loss to supervise the prediction of scene categories, as defined by the following equation:
\begin{equation}
\mathcal{L}_{\mathrm{scene}} = -\frac{1}{B} \sum_{i=1}^B \sum_{c=0}^1 y_{i, c} \log \left(p_{i, c}\right)
\end{equation}
where $y_{i, c}$ denotes the one-hot encoded ground truth label of the scene category derived from the radar point cloud, $p_{i, c}$ is the predicted probability for the scene category, and $B$ represents the batch size.
To determine when to terminate training, we use a Mean Squared Error (MSE)-based evaluation criterion defined as follows: 
\begin{equation}
\mathcal{L}_{\mathrm{MSE}}=\frac{1}{B} \sum_{i=1}^B\left(\hat{n}_i-n_i\right)^2
\end{equation}
where $\hat{n}_i,\; n_i \in \{1, 2\}$ represent the ground truth and predicted scene categories, respectively. A value of 1 indicates a scene containing only buildings or trees, while a value of 2 indicates a scene with both buildings and trees. The ground truth label $\hat{n}_i$ is obtained directly from the label, whereas the predicted label $n_i$ is inferred from the class probabilities $p_{i, c}$.
Training is terminated if both of the following conditions are satisfied:
\begin{enumerate}
    \item At least 90\% of the validation samples achieve a local MSE (LMSE) below 0.1.
    \item The validation loss fluctuates by less than 5\% over five consecutive epochs.
\end{enumerate}
In the first stage, we train the network using $p_{i, c}$ to ensure proper gradient updates, while using $n_i$ for termination criteria. This design guaranties that the training process remains stable while maintaining the accuracy of the outputs fed into subsequent prediction stages.
\begin{figure}[t]
\centering
\subfigure[]{\includegraphics[width=3.3in]{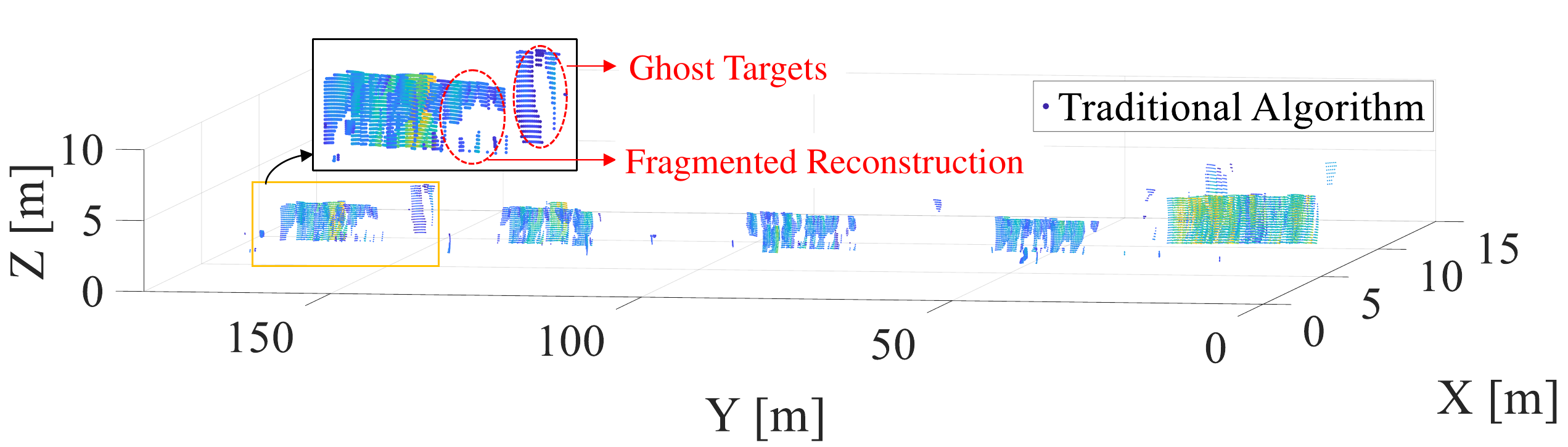}}
\subfigure[]{\includegraphics[width=3.3in]{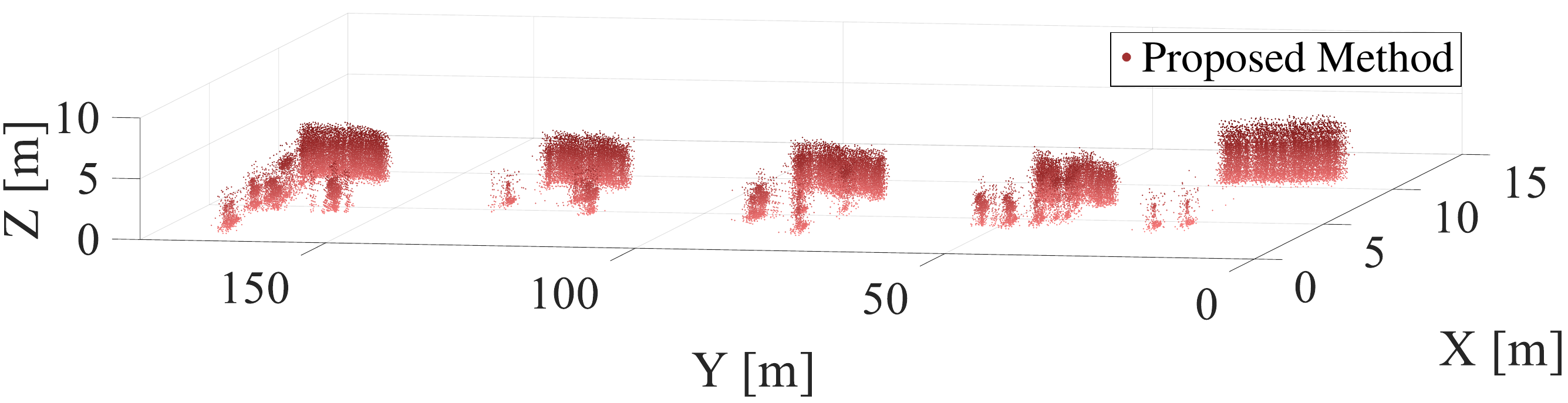}}
\subfigure[]{\includegraphics[width=3.3in]{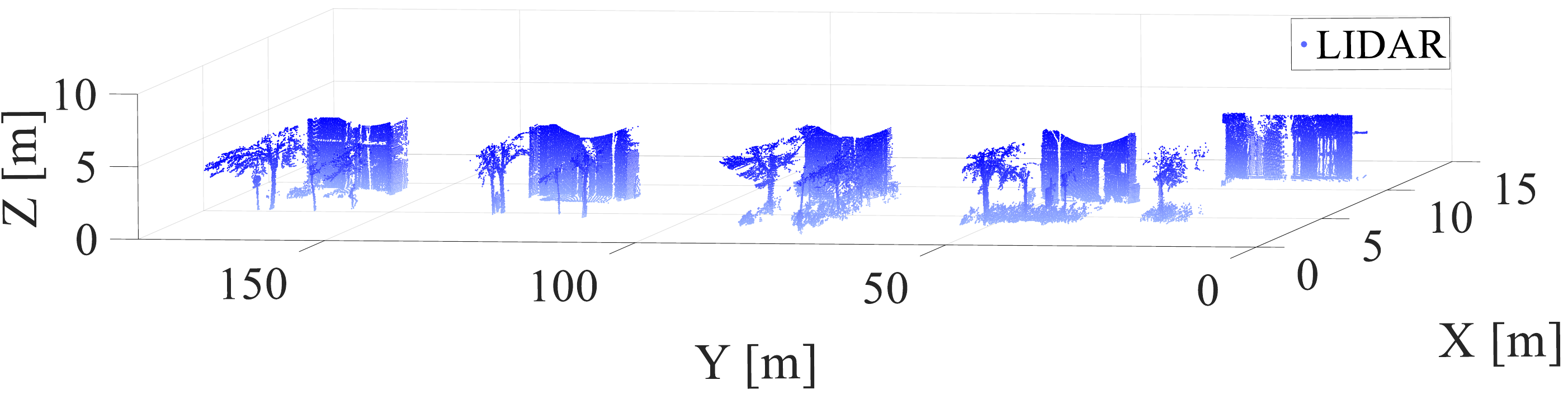}}
\caption{Comparison of proposed method with traditional method. (a) Reconstructing scenarios with traditional methods. (b) Reconstructing scenarios with proposed method. (c) LiDAR point cloud.}
\label{fig:13}
\end{figure}
\subsubsection{Training Stage II: Semantic center position}
We use MSE to measure the discrepancy between the predicted and ground-truth semantic object centers. The corresponding loss function is defined as follows:
\begin{equation}
\mathcal{L}_{\text {center }}=\frac{1}{B} \sum_{i=1}^B \frac{1}{K} \sum_{k=1}^K\left\|\hat{C}_{i, \sigma(k)}-C_{i, \sigma(k)}\right\|_2^2
\end{equation}
% where $\hat{C}_i \in \mathbb{R}^{K \times 3}$ denotes the predicted coordinates of the cluster centers ($K=2$ being the maximum number of clusters), and $C_i$ are the coordinates of the center of the ground truth cluster. The function $\sigma(\cdot)$ represents a sorting index based on ascending x coordinate values. For scenes that contain both buildings and trees, this sorting step ensures a consistent correspondence between predicted and ground-truth cluster centers, enabling correct matching between the respective building and tree clusters.
where $\hat{C}_{i}\in\mathbb{R}^{K\times3}$ denotes the predicted coordinates of the semantic object centers, and $C_{i}$ represent the ground-truth centers, derived by calculating the geometric centroid of all LiDAR points belonging to a specific semantic category (manually annotated as Tree or Building). We set the maximum number of semantic categories to $K=2$, covering the dominant scatterer types in the dataset. The function $\sigma(\cdot)$ sorts the centers based on ascending $x$-coordinate values. This sorting strategy is physically motivated by the typical urban street layout, where roadside trees are positioned closer to the sensing vehicle (smaller $x$-value) than the building facades behind them. This ensures a consistent one-to-one correspondence between the predicted and ground-truth centers during the loss calculation for mixed scenes.
This stage of training is terminated when the average $\mathcal{L}_{\text {center }}$ in the validation set falls below 2.0.

\subsubsection{Training Stage III: Point cloud details}
In the point cloud generation stage, we adopt an adaptively weighted multi-objective loss function:
\begin{equation}
\mathcal{L}_{\mathrm{point cloud}}=\sum_{i=1}^2 (\exp \left(-\lambda_i\right) \mathcal{L}_i+\lambda_i)+\mathcal{L}_{\text {VAT }}
\end{equation}

where $\lambda_i$ is a trainable log-variance parameter. $\mathcal{L}_{\mathrm{point cloud}}$ consists of two main components: 
\begin{enumerate}
    \item $Chamfer Distance (CD)$ \cite{wu2021balanced}, which restricts the global geometric structure.
    \begin{equation}
    \begin{aligned}
      \mathcal{L}_{\mathrm{1}} = \frac{1}{B N} \sum_{i=1}^B \Bigg(
      &\frac{1}{\left|P_i\right|} \sum_{p \in P_i} \min_{q \in Q_i} \|p - q\|_2 \\
      + &\frac{1}{\left|Q_i\right|} \sum_{q \in Q_i} \min_{p \in P_i} \|q - p\|_2
      \Bigg)
    \end{aligned}
    \end{equation}
    Specifically, let $P_i$ and $Q_i$ denote the predicted and ground truth point clouds, respectively, with $N=1000$ points.
    \item $Local Feature Loss$ \cite{feng2019point2sequence}, which captures fine-grained details of the predicted point cloud.
    \begin{equation}
    \mathcal{L}_{\text {2}}=\frac{1}{B} \sum_{i=1}^B \frac{1}{k} \sum_{j=1}^k\left\|\mu\left(\mathcal{N}_j^P\right)-\mu\left(\mathcal{N}_j^Q\right)\right\|_2
    \end{equation}
    where $\mathcal{N}_j$ denotes the set of 50 nearest neighbors of point $j$, and $\mu(\cdot)$ computes the centroid of the local neighborhood.
\end{enumerate}
\begin{table}[t]
\centering
\caption{Quantitative comparison of MSCR-Net performance with different numbers of input MPCs.}
\label{tab:mpc_ablation}
\renewcommand{\arraystretch}{1.2}
\begin{tabular}{c|cc}
\hline
\textbf{Number of MPCs} & \textbf{CD [m]} $\downarrow$ & \textbf{F-score@0.5m} $\uparrow$ \\ \hline
20 & 0.5973 & 0.9095 \\
40 & 0.5846 & 0.9015 \\
60 & 0.6063 & 0.8925 \\ \hline
\end{tabular}
\end{table}
% For training at this stage, we use the Adam optimizer with an initial learning rate of 0.0001 and a weight decay of 0.0001. The batch size is set to 10. In addition, we incorporate Virtual Adversarial Training (VAT) \cite{miyato2018virtual} to enhance the robustness of the model.
Regarding the optimization details, we employ the Adam optimizer with a learning rate of $1 \times 10^{-4}$ across all three stages. Crucially, the optimizer is re-initialized at the beginning of each stage: Stage I optimizes the scene predictor and encoders; Stage II optimizes the semantic center decoder while freezing the encoders; and Stage III optimizes the point cloud decoder while freezing all preceding modules. The batch size is set to 10, for which the variable-length cropped point clouds are uniformly downsampled to a fixed size of $N=1000$ to facilitate batch processing. In addition, we incorporate Virtual Adversarial Training (VAT) \cite{miyato2018virtual} to enhance the robustness of the model. The VAT is configured with a loss weight $\alpha=1.0$, a perturbation magnitude $\epsilon=10^{-3}$, an initial variance $\xi=10^{-5}$, and employs an $L_2$-norm constraint with a single power iteration step.

\section{EXPERIMENTS AND ANALYSIS}
% In this section, we first introduce the evaluation metrics used to assess the performance of proposed MSCR-Net. Subsequently, we conducted experiments on the dataset described in Section IV. To the best of our knowledge, there has been no prior work that leverages deep learning to reconstruct environment based on ISAC channel data. Therefore, we evaluated the effectiveness of MSCR-Net through a series of experiments, including ablation studies and spatial consistency analysis. Specifically, the ablation studies focus on the analysis of local features in sample-level reconstruction results, while the spatial consistency experiment evaluates the dynamic global characteristics of the reconstructed environment.
In this section, we first introduce the evaluation metrics used to assess the performance of the proposed MSCR-Net. Subsequently, we present the experiments conducted on the dataset described in Section IV. To comprehensively evaluate the effectiveness of MSCR-Net, we designed a series of experiments comprising both comparative evaluations against state-of-the-art models and ablation studies. Specifically, comparative experiments benchmark our method against traditional algorithms and advanced deep learning baselines to demonstrate its performance superiority. Ablation studies focus on analyzing the contribution of specific modules to the local features in sample-level reconstruction, while the spatial consistency experiment evaluates the dynamic global characteristics of the reconstructed environment.
\begin{table}[t]
\centering
\caption{Performance and Complexity Comparison}
\begin{tabular}{lcccc}
\toprule
\textbf{Methods} & \textbf{CD ↓} & \textbf{F-score@0.5m ↑} & \textbf{Elevation} & \textbf{Complexity} \\
\midrule
MSCR-Net            & 0.29\,m  & 0.87  & 7.97\,dB & 0.9\,ms \\
Traditional & 1.81\,m  & 0.38  & 3.53\,dB & 15\,ms \\
LiDAR               & 0\,m     & 1.00 & 0.00\,dB & 100\,ms \\
\bottomrule
\end{tabular}
\label{tab:elevation_results}
\end{table}
\subsection{Evaluation Metrics}
Let $\mathcal{P}_{\text {out }}$ denote the point cloud reconstruction, and let $\mathcal{P}_{G T}$ represent the set of LiDAR point clouds. The evaluation employs two metrics:
\begin{enumerate}
    \item $Chamfer Distance (CD)$ measures the closeness between two sets of points based on nearest neighbor distances. In the computation, it only needs to replace $P_i$ in (8) with $\mathcal{P}_{\text {out }}$, $Q_i$ with $\mathcal{P}_{G T}$, and set $B=1$.
    \item $F-score@0.5m$ \cite{tatarchenko2019single} assesses the quality of reconstruction through precision and recall with a distance threshold $tr=0.5m$. This threshold is chosen to enforce strict geometric accuracy relative to the large spatial scale of dynamic vehicular environments (spanning tens of meters):
    \begin{equation}
    F\text{-score}\left(\mathcal{P}_{\text{out}}, \mathcal{P}_{GT}\right)=\frac{2 \times \mathcal{P}_{tr} \times R_{tr}}{\mathcal{P}_{tr}+R_{tr}}
    \end{equation}
    where $\mathcal{P}_{tr}$ and $R_{tr}$ denote precision and recall, respectively:
    \begin{equation}
    \begin{gathered}
    \mathcal{P}_{tr} = \frac{1}{\left|\mathcal{P}_{\text{out}}\right|} \sum_{x \in \mathcal{P}_{\text{out}}} \left[\textit{min}_{y \in \mathcal{P}_{GT}} \|x-y\|_{2}<tr\right] \\
    R_{tr} = \frac{1}{\left|\mathcal{P}_{GT}\right|} \sum_{x \in \mathcal{P}_{GT}} \left[\textit{min}_{y \in \mathcal{P}_{\text{out}}} \|x-y\|_{2}<tr\right]
    \end{gathered}
    \end{equation}
\end{enumerate}
It is apparent that smaller values of CD correspond to higher accuracy in reconstruction, while larger values of F-score@0.5m indicate a wider coverage range in reconstruction.
% \begin{figure}[t]
%   \centering

%   \begin{subfigure}[b]{0.48\textwidth}
%     \centering
%     \includegraphics[width=\linewidth]{result-predandlidarv2.pdf}
%     \caption{Comparison of reconstructed scene and LiDAR ground truth in a unified coordinate system.}
%     \label{fig:combined}
%   \end{subfigure}

%   \begin{subfigure}[b]{0.48\textwidth}
%     \centering
%     \includegraphics[width=\linewidth]{result-predandlidar-xy-v2.pdf}
%     \caption{The xy-plane of Fig. \ref{fig:12}(a).}
%     \label{fig:xyplane}
%   \end{subfigure}

%   \caption{Comparison of proposed method with LiDAR.}
%   \label{fig:12}
% \end{figure}

\subsection{Performance Comparison and Ablation Studies}

We benchmarked MSCR-Net against two distinct baseline strategies: 1) ResNet \cite{xu2021indoor}, a standard deep network that treats channel parameters as three-channel images, ignoring the specific physical properties of ISAC signals; and 2) AGGRNN \cite{10636720}, a two-step reconstruction framework that first maps channel parameters to sparse point clouds in 3D space and subsequently utilizes a densification network to generate the final environment.
As visually demonstrated in Fig. \ref{fig:result-singlev2}, the ResNet baseline exhibits significant inaccuracies. While it can approximately locate structures in single-object scenes, it fails to identify scenes with multiple types of scatterers and cannot distinguish between buildings and trees, resulting in blurred and fused outlines. AGGRNN shows improved performance over ResNet by explicitly utilizing spatial point representations. However, its reconstruction quality is limited by the sparsity of the intermediate representation, leading to loss of fine structural details and geometric distortions in complex mixed scenes.
In contrast, MSCR-Net achieves the most complete scene reconstructions. By using the specific physics of the channel data, it generates sharper outlines and accurately distinguishes between buildings and trees as distinct scatterers, resulting in high consistency with the LiDAR ground truth.

To investigate the necessity of the proposed multi-stage architecture, we evaluated two ablation variants: 1) E2E-Transformer, a single-stage baseline that employs a Transformer-based architecture to directly map channel parameters to point clouds in an end-to-end manner; and 2) CR-Net (Channel-to-Reconstruction Network), which retains the encoder and point cloud decoder of MSCR-Net but removes the progressive multi-stage training structure, effectively stacking the decoders and supervising them solely by the final reconstruction loss.
The results reveal that the E2E-Transformer, despite its powerful attention mechanism, struggles to converge to a precise geometric structure due to the huge modality gap between abstract channel features and dense point clouds. The CR-Net shows better spatial localization than the single-stage Transformer but remains indistinct in boundary definition. As seen in mixed scenes (Fig. \ref{fig:result-singlev2}), CR-Net tends to merge proximal trees and buildings into a single object. This confirms that without the progressive guidance of scene classification (Stage I) and semantic center Positioning (Stage II), the network struggles to decouple the complex spatial relationships.

Figs. \ref{fig:CD-1} and \ref{fig:CD-2} present the quantitative performance of all five networks on the test set. MSCR-Net outperforms both the comparative baselines (ResNet, AGGRNN) and the ablation variants (E2E-Transformer, CR-Net) across all evaluation metrics. It demonstrates a superior capability in reconstructing point clouds with richer local and global structural features. Furthermore, MSCR-Net exhibits the smallest dispersion in both CD and F-score@0.5m, whereas the baselines show significantly higher variance. This highlights that MSCR-Net possesses strong generalization ability and robustly handles the non-linear mapping from ISAC channels to environmental structures.

Finally, regarding the robustness to the number of MPCs, we conducted additional ablation studies with MPC counts of 20 and 60, comparing them to our default setting of 40. As summarized in Table II, the 40-MPC setting achieves the optimal geometric accuracy with the lowest Average CD. In comparison, the 20-MPC configuration yields a slightly higher geometric error, indicating potential information loss from insufficient paths. In contrast, increasing to 60 MPCs leads to performance degradation, likely due to the introduction of noise from weaker signal components. Despite these minor fluctuations, the overall performance variance is minimal, demonstrating that MSCR-Net is robust to variations in input sparsity.

\subsection{Spatial Consistency}
In dynamic traffic environments, although the acquisition rate of ISAC channel data is higher than the rate of scene variation, the continuous motion of the vehicle leads to changes in both the observation angle and spatial position associated with each channel snapshot. Consequently, even when multi-ple samples perceive the same scene, the extracted channel features exhibit differences. These differences not only reflect variations in viewing angles across samples, but also capture the dynamic nature of channel perception with respect to spatial structures. The model’s ability to effectively capture and leverage these variations is critical for achieving spatially consistent reconstruction. It directly influences the completeness and stability of dynamic 3D scene reconstruction, helping to prevent perception misalignment and the emergence of ghost targets \cite{zheng2024detection}.

% We performed temporal reconstruction on the validation set samples by cumulatively mapping reconstructed point clouds along the y-axis according to the vehicle’s measured speed and the channel sampling rate, thus restoring the perceived scene of the channel over time. Fig. \ref{fig:13}(a) shows the global reconstruction results of the traditional method \cite{song20243d}, and it can be found that many buildings are incomplete, the trees are poorly reconstructed, and there are ghost targets.
We performed temporal reconstruction on the test set samples by cumulatively mapping the reconstructed point clouds along the y-axis according to the vehicle’s measured speed and the channel sampling rate. To provide a performance benchmark, we evaluated a traditional geometry-based channel inversion method \cite{song20243d}. This baseline combines Digital Beamforming (DBF) to scan for angular directions using 2D steering vectors and Matched Filtering (MF) to estimate target distances based on the peak offsets of the received echo correlation. Fig. \ref{fig:13}(a) illustrates the global reconstruction results of this traditional approach. It can be observed that the reconstruction quality is suboptimal: many building structures are incomplete, trees are poorly resolved, and significant ghost targets are present due to multipath interference.

% Notably, the unstable spatial structure of trees leads traditional approaches to frequently produce ghosting effects around tree clusters.

In contrast, Fig. \ref{fig:12} and Fig. \ref{fig:13}(b) present the global reconstruction using the proposed MSCR-Net. The reconstructed geometry changes smoothly over time, maintaining high continuity with the true environment, and crucially, no ghost targets are observed around trees or other scatterers. It is important to note that in the test scenarios, there are almost no effective scatterers between adjacent buildings that can be captured by the sensing system. As a result, such samples were excluded during dataset construction. Therefore, the reconstructed environments shown in Fig. \ref{fig:12} and Fig. \ref{fig:13} are not continuous.
TABLE \ref{tab:elevation_results} compares the performance of the proposed MSCR-Net with that of traditional methods. It is important to note that unlike the single-snapshot evaluation in Figs. 8 and 9, the metrics in TABLE \ref{tab:elevation_results} are calculated based on the complete, globally reconstructed environment (Fig. 11(b)). The lower CD (0.29) achieved here compared to the single-snapshot result (0.5846) is due to the accumulation of overlapping snapshots, which densifies the point cloud and improves structural completeness. MSCR-Net outperforms traditional approaches across all evaluation metrics, demonstrating superior reconstruction accuracy. The column 'Elevation (dB)' indicates the relative improvement in reconstruction quality compared to the traditional baseline. This demonstrates that the model tracks the evolution of the environment in time well and is more suitable for dynamic environment reconstruction. Note that since LiDAR's method is treated as the true value of the benchmark in this work, the metrics here are of no practical significance and are listed here to analyze their complexity below.
\subsection{Complexity Analysis}
To evaluate the feasibility of the proposed method for real-world deployment, we analyzed the model's parameter count, computational complexity, and inference speed. The overall parameter size of MSCR-Net is 15.53M, with a computational complexity of approximately 0.53 GFLOPs, demonstrating high inference efficiency while maintaining strong reconstruction performance. On an NVIDIA RTX 5080 GPU, the proposed method achieves an average inference time of approximately 0.9 ms per sample, significantly outperforming traditional reconstruction algorithms (15 ms) and LiDAR-based methods (100 ms). Even on a general-purpose computing platform without GPU acceleration (using an Intel Core Ultra 9 275HX CPU with 24 cores and a base speed of 2.70 GHz), MSCR-Net achieves an inference time of only 3 ms per sample, further highlighting its strong hardware adaptability and potential for edge deployment.

Finally, we address the generalization capability regarding hardware changes and new scenes. While real-world data collection is labor-intensive, the proposed architecture captures fundamental physical mappings between channel parameters and spatial structures, which are independent of specific hardware. Therefore, when hardware configurations (e.g., antenna placement or frequency) change, transfer learning can be effectively employed to fine-tune the pre-trained backbone with minimal new data, significantly reducing recollection overhead \cite{omondi2026advancing,hu2024transfer}. Furthermore, since the model learns the consistent electromagnetic scattering properties of typical urban elements (buildings and trees), it has strong potential to generalize across other urban vehicular environments with similar structural compositions.

\section{Conclusion}
This paper presents an ISAC-driven reconstruction framework tailored for dynamic vehicular environment, which jointly leverages mmWave ISAC channel parameters and LiDAR-derived point cloud data to achieve efficient environment reconstruction. Experimental results demonstrate that in complex dynamic scenes composed of diverse scatterers, the proposed model significantly outperforms the traditional beamforming-based methods and deep learning baselines, achieving a Chamfer Distance (CD) of 0.29 and an F-score@0.5m of 0.87. This study further verifies that, when properly extracted and modeled, channel characteristics can serve as a powerful alternative to radar sensing, offering a low-cost, highly integrated solution for 6G V2X applications. Future work will focus on extending the proposed framework to multi-vehicle cooperative sensing tasks and exploring its real-time deployment potential on edge computing platforms.

\bibliographystyle{IEEEtran}
\bibliography{text}

@article{wei2023integrated,
  title={Integrated sensing and communication signals toward {5G-A} and {6G}: A survey},
  author={Wei, Zhiqing and Qu, Hanyang and Wang, Yuan and Yuan, Xin and Wu, Huici and Du, Ying and Han, Kaifeng and Zhang, Ning and Feng, Zhiyong},
  journal={IEEE Internet of Things Journal},
  volume={10},
  number={13},
  pages={11068--11092},
  year={2023},
  publisher={IEEE}
}

@article{sun2020mimo,
  title={{MIMO} radar for advanced driver-assistance systems and autonomous driving: Advantages and challenges},
  author={Sun, Shunqiao and Petropulu, Athina P and Poor, H Vincent},
  journal={IEEE Signal Processing Magazine},
  volume={37},
  number={4},
  pages={98--117},
  year={2020},
  publisher={IEEE}
}

@article{qiu2025multi,
  title={A Multi-Feature Fusion Approach for Road Surface Recognition Leveraging Millimeter-Wave Radar},
  author={Qiu, Zhimin and Shao, Jinju and Guo, Dong and Yin, Xuehao and Zhai, Zhipeng and Duan, Zhibing and Xu, Yi},
  journal={Sensors},
  volume={25},
  number={12},
  pages={3802},
  year={2025},
  publisher={MDPI}
}

@article{zifeng20254d,
  title={{4D} Millimeter-Wave Radar Target Detection Method Based on Graph Neural Network.},
  author={Zifeng, HUANG and Hongyan, WANG and Jiakang, MA},
  journal={Journal of Computer Engineering \& Applications},
  volume={61},
  number={10},
  year={2025}
}

@article{luo2025improving,
  title={Improving Multi-Vehicle Perception Fusion with Millimeter-Wave Radar Assistance},
  author={Luo, Zhiqing and Wang, Yi and He, Yingying and Wang, Wei},
  journal={arXiv preprint arXiv:2506.00837},
  year={2025}
}

@article{chen2024enhancing,
  title={Enhancing Integrated Sensing and Communication ({ISAC}) Performance for a Searching--Deciding Alternation Radar-Comm System with Multi-Dimension Point Cloud Data},
  author={Chen, Leyan and Liu, Kai and Gao, Qiang and Wang, Xiangfen and Zhang, Zhibo},
  journal={Remote Sensing},
  volume={16},
  number={17},
  pages={3242},
  year={2024},
  publisher={MDPI}
}

@article{chen2025rageosense,
  title={RaGeoSense for smart home gesture recognition using sparse millimeter wave radar point clouds},
  author={Chen, Honghong and Wang, Xiangyu and Hao, Zhanjun and Lu, Yuru and Li, Jingyu and Zhang, Haozhe and Xi, Ben},
  journal={Scientific Reports},
  volume={15},
  number={1},
  pages={15267},
  year={2025},
  publisher={Nature Publishing Group UK London}
}

@ARTICLE{cheng2023m,
  author={Cheng, Xiang and Huang, Ziwei and Bai, Lu and Zhang, Haotian and Sun, Mingran and Liu, Boxun and Li, Sijiang and Zhang, Jianan and Lee, Minson},
  journal={China Communications}, 
  title={{M3SC}: A generic dataset for mixed multi-modal ({MMM}) sensing and communication integration}, 
  year={2023},
  volume={20},
  number={11},
  pages={13-29},
  doi={10.23919/JCC.fa.2023-0268.202311}}

@article{blandino2025detecting,
  title={Detecting Airborne Objects with {5G NR} Radars},
  author={Blandino, Steve and Golmie, Nada and Sahoo, Anirudha and Nguyen, Thao and Ropitault, Tanguy and Griffith, David and Sonny, Amala},
  journal={arXiv preprint arXiv:2505.24763},
  year={2025}
}

@article{chang2024environment,
  title={Environment Reconstruction with Multi-targets Reflectors-merged Sensing Method Based on {THz} Single-sided Channel Characteristics},
  author={Chang, Zhaowei and Tang, Pan and Zhang, Jianhua and Jiang, Hao and Liu, Guangyi},
  journal={arXiv preprint arXiv:2412.03960},
  year={2024}
}

@article{avazov2021trajectory,
  title={A trajectory-driven {3D} non-stationary mm-wave {MIMO} channel model for a single moving point scatterer},
  author={Avazov, Nurilla and Hicheri, Rym and Muaaz, Muhammad and Sanfilippo, Filippo and P{\"a}tzold, Matthias},
  journal={IEEE Access},
  volume={9},
  pages={115990--116001},
  year={2021},
  publisher={IEEE}
}

@article{du2022integrated,
  title={Integrated sensing and communications for {V2I} networks: Dynamic predictive beamforming for extended vehicle targets},
  author={Du, Zhen and Liu, Fan and Yuan, Weijie and Masouros, Christos and Zhang, Zenghui and Xia, Shuqiang and Caire, Giuseppe},
  journal={IEEE Transactions on Wireless Communications},
  volume={22},
  number={6},
  pages={3612--3627},
  year={2022},
  publisher={IEEE}
}

@inproceedings{song20243d,
  title={{3D} Environment Reconstruction Based on {ISAC} Channels},
  author={Song, Junzhe and He, Ruisi and Zhang, Zhengyu and Yang, Mi and Ai, Bo and Zhang, Haoxiang and Chen, Ruifeng},
  booktitle={2024 International Conference on Ubiquitous Communication (Ucom)},
  pages={487--491},
  year={2024},
  organization={IEEE}
}

@inproceedings{lu2020see,
  title={See through smoke: robust indoor mapping with low-cost {mmWave} radar},
  author={Lu, Chris Xiaoxuan and Rosa, Stefano and Zhao, Peijun and Wang, Bing and Chen, Changhao and Stankovic, John A and Trigoni, Niki and Markham, Andrew},
  booktitle={Proceedings of the 18th International Conference on Mobile Systems, Applications, and Services},
  pages={14--27},
  year={2020}
}

@inproceedings{qi2017pointnet,
  title={{Pointnet}: Deep learning on point sets for {3D} classification and segmentation},
  author={Qi, Charles R and Su, Hao and Mo, Kaichun and Guibas, Leonidas J},
  booktitle={Proceedings of the IEEE conference on computer vision and pattern recognition},
  pages={652--660},
  year={2017}
}

@inproceedings{fan2017point,
  title={A point set generation network for {3D} object reconstruction from a single image},
  author={Fan, Haoqiang and Su, Hao and Guibas, Leonidas J},
  booktitle={Proceedings of the IEEE conference on computer vision and pattern recognition},
  pages={605--613},
  year={2017}
}

@article{qi2017pointnet++,
  title={{Pointnet++}: Deep hierarchical feature learning on point sets in a metric space},
  author={Qi, Charles Ruizhongtai and Yi, Li and Su, Hao and Guibas, Leonidas J},
  journal={Advances in neural information processing systems},
  volume={30},
  year={2017}
}

@article{wu2021balanced,
  title={Balanced chamfer distance as a comprehensive metric for point cloud completion},
  author={Wu, Tong and Pan, Liang and Zhang, Junzhe a   nd Wang, Tai and Liu, Ziwei and Lin, Dahua},
  journal={Advances in Neural Information Processing Systems},
  volume={34},
  pages={29088--29100},
  year={2021}
}

@inproceedings{tatarchenko2019single,
  title={What do single-view {3D} reconstruction networks learn?},
  author={Tatarchenko, Maxim and Richter, Stephan R and Ranftl, Ren{\'e} and Li, Zhuwen and Koltun, Vladlen and Brox, Thomas},
  booktitle={Proceedings of the IEEE/CVF conference on computer vision and pattern recognition},
  pages={3405--3414},
  year={2019}
}

@inproceedings{feng2019point2sequence,
  title={{Point2Sequence}: Learning the shape representation of {3D} point clouds with an attention-based sequence to sequence network},
  author={Feng, Yifan and Zhang, Yao and Zhao, Zizhao and Ji, Rongrong and Gao, Yue},
  booktitle={Proceedings of the AAAI Conference on Artificial Intelligence},
  volume={33},
  pages={8439--8446},
  year={2019}
}

@article{lu2024integrated,
  title={Integrated sensing and communications: Recent advances and ten open challenges},
  author={Lu, Shihang and Liu, Fan and Li, Yunxin and Zhang, Kecheng and Huang, Hongjia and Zou, Jiaqi and Li, Xinyu and Dong, Yuxiang and Dong, Fuwang and Zhu, Jia and others},
  journal={IEEE Internet of Things Journal},
  volume={11},
  number={11},
  pages={19094--19120},
  year={2024},
  publisher={IEEE}
}

@article{zhang2023integrated,
  title={Integrated sensing and communication channel: Measurements, characteristics, and modeling},
  author={Zhang, Jianhua and Wang, Jialin and Zhang, Yuxiang and Liu, Yameng and Chai, Zeyong and Liu, Guangyi and Jiang, Tao},
  journal={IEEE Communications Magazine},
  volume={62},
  number={6},
  pages={98--104},
  year={2023},
  publisher={IEEE}
}

@article{katoh2025multidlformer,
  title={{MultiDLFormer}: A Vision {Transformer-Based} Deep Learning Model for Detecting Deep {Low-Frequency} Earthquakes Across Multiple Stations},
  author={Katoh, Shinya and Nagao, Hiromichi and Iio, Yoshihisa},
  journal={Authorea Preprints},
  year={2025},
  publisher={Authorea}
}

@article{yun2025toast,
  title={{TOAST}: Task-Oriented Adaptive Semantic Transmission over Dynamic Wireless Environments},
  author={Yun, Sheng and Pei, Jianhua and Wang, Ping},
  journal={arXiv preprint arXiv:2506.21900},
  year={2025}
}

@article{liu2025ai,
  title={{AI}-Empowered Channel Generation for {IoV} Semantic Communications in Dynamic Conditions},
  author={Liu, Hao and Yang, Bo and Yu, Zhiwen and Cao, Xuelin and Alexandropoulos, George C and Zhang, Yan and Yuen, Chau},
  journal={arXiv preprint arXiv:2507.02013},
  year={2025}
}

@inproceedings{klautau2021generating,
  title={Generating {MIMO} channels for {6G} virtual worlds using ray-tracing simulations},
  author={Klautau, Aldebaro and de Oliveira, Ailton and Trindade, Isabela Pamplona and Alves, Wesin},
  booktitle={2021 IEEE Statistical Signal Processing Workshop (SSP)},
  pages={595--599},
  year={2021},
  organization={IEEE}
}

@article{zheng2024detection,
  title={Detection of ghost targets for automotive radar in the presence of multipath},
  author={Zheng, Le and Long, Jiamin and Lops, Marco and Liu, Fan and Hu, Xueyao and Zhao, Chuanhao},
  journal={IEEE Transactions on Signal Processing},
  year={2024},
  publisher={IEEE}
}

@article{miyato2018virtual,
  title={Virtual adversarial training: a regularization method for supervised and semi-supervised learning},
  author={Miyato, Takeru and Maeda, Shin-ichi and Koyama, Masanori and Ishii, Shin},
  journal={IEEE transactions on pattern analysis and machine intelligence},
  volume={41},
  number={8},
  pages={1979--1993},
  year={2018},
  publisher={IEEE}
}

@article{fessler2002space,
  title={Space-alternating generalized expectation-maximization algorithm},
  author={Fessler, Jeffrey A and Hero, Alfred O},
  journal={IEEE Transactions on signal processing},
  volume={42},
  number={10},
  pages={2664--2677},
  year={2002},
  publisher={IEEE}
}

@inproceedings{chong2002joint,
  title={Joint detection-estimation of directional channel parameters using the {2-D} frequency domain {SAGE} algorithm with serial interference cancellation},
  author={Chong, Chia-Chin and Laurenson, David I and Tan, Chor Min and McLaughlin, Steve and Beach, Mark A and Nix, Andrew R},
  booktitle={2002 IEEE International Conference on Communications. Conference Proceedings. ICC 2002 (Cat. No. 02CH37333)},
  volume={2},
  pages={906--910},
  year={2002},
  organization={IEEE}
}

@inproceedings{matthaiou2007characterization,
  title={Characterization of an indoor {MIMO} channel in frequency domain using the {3D-SAGE} algorithm},
  author={Matthaiou, Michail and Laurenson, David I and Razavi-Ghods, Nima and Salous, Sana},
  booktitle={2007 IEEE International Conference on Communications},
  pages={5868--5872},
  year={2007},
  organization={IEEE}
}

@article{hotelling1933analysis,
  title={Analysis of a complex of statistical variables into principal components.},
  author={Hotelling, Harold},
  journal={Journal of educational psychology},
  volume={24},
  number={6},
  pages={417},
  year={1933},
  publisher={Warwick \& York}
}

@article{shlens2014tutorial,
  title={A tutorial on principal component analysis},
  author={Shlens, Jonathon},
  journal={arXiv preprint arXiv:1404.1100},
  year={2014}
}

@article{he2019propagation,
  title={Propagation channels of {5G} millimeter-wave vehicle-to-vehicle communications: Recent advances and future challenges},
  author={He, Ruisi and Schneider, Christian and Ai, Bo and Wang, Gongpu and Zhong, Zhangdui and Dupleich, Diego A and Thomae, Reiner S and Boban, Mate and Luo, Jian and Zhang, Yunyong},
  journal={IEEE vehicular technology magazine},
  volume={15},
  number={1},
  pages={16--26},
  year={2019},
  publisher={IEEE}
}

@article{huang2022artificial,
  title={Artificial intelligence enabled radio propagation for communications—Part {II}: Scenario identification and channel modeling},
  author={Huang, Chen and He, Ruisi and Ai, Bo and Molisch, Andreas F and Lau, Buon Kiong and Haneda, Katsuyuki and Liu, Bo and Wang, Cheng-Xiang and Yang, Mi and Oestges, Claude and others},
  journal={IEEE Transactions on Antennas and Propagation},
  volume={70},
  number={6},
  pages={3955--3969},
  year={2022},
  publisher={IEEE}
}

@article{he2017geometrical,
  title={Geometrical-based modeling for millimeter-wave {MIMO} mobile-to-mobile channels},
  author={He, Ruisi and Ai, Bo and St{\"u}ber, Gordon L and Wang, Gongpu and Zhong, Zhangdui},
  journal={IEEE transactions on vehicular technology},
  volume={67},
  number={4},
  pages={2848--2863},
  year={2017},
  publisher={IEEE}
}

@book{he2024wireless,
  title={Wireless channel measurement and modeling in mobile communication scenario: Theory and application},
  author={He, Ruisi and Ai, Bo},
  year={2024},
  publisher={CRC press}
}

@article{zhang2022general,
  title={A general channel model for integrated sensing and communication scenarios},
  author={Zhang, Zhengyu and He, Ruisi and Ai, Bo and Yang, Mi and Li, Chao and Mi, Hang and Zhang, Zhangdui},
  journal={IEEE Communications Magazine},
  volume={61},
  number={5},
  pages={68--74},
  year={2022},
  publisher={IEEE}
}

@article{zhang2024cluster,
  title={A cluster-based statistical channel model for integrated sensing and communication channels},
  author={Zhang, Zhengyu and He, Ruisi and Ai, Bo and Yang, Mi and Niu, Yong and Zhong, Zhangdui and Li, Yujian and Zhang, Xuejian and Li, Jing},
  journal={IEEE Transactions on Wireless Communications},
  volume={23},
  number={9},
  pages={11597--11611},
  year={2024},
  publisher={IEEE}
}

@article{tian2025analytical,
  title={Analytical channel modeling: From {MIMO} to extra large-scale {MIMO}},
  author={Tian, Jiachen and Han, Yu and Jin, Shi and Zhang, Jun and Wang, Jue},
  journal={Chinese Journal of Electronics},
  volume={34},
  number={1},
  pages={1--15},
  year={2025},
  publisher={CIE}
}

@article{xu2021indoor,
  title={An indoor localization system using residual learning with channel state information},
  author={Xu, Chendong and Wang, Weigang and Zhang, Yunwei and Qin, Jie and Yu, Shujuan and Zhang, Yun},
  journal={Entropy},
  volume={23},
  number={5},
  pages={574},
  year={2021},
  publisher={MDPI}
}

@ARTICLE{10839242,
  author={He, Ruisi and Cicco, Nicola D. and Ai, Bo and Yang, Mi and Miao, Yang and Boban, Mate},
  journal={IEEE Wireless Communications}, 
  title={{COST CA20120} INTERACT Framework of Artificial Intelligence-Based Channel Modeling}, 
  year={2025},
  volume={32},
  number={4},
  pages={200-207},
  doi={10.1109/MWC.010.2400253}}

@ARTICLE{10644121,
  author={He, Boxiang and Wang, Fanggang and Quek, Tony Q. S.},
  journal={IEEE Wireless Communications Letters}, 
  title={Secure Semantic Communication via Paired Adversarial Residual Networks}, 
  year={2024},
  volume={13},
  number={10},
  pages={2832-2836},
  doi={10.1109/LWC.2024.3448474}}

@ARTICLE{10945996,
  author={Yin, Lannuo and Wang, Yong},
  journal={IEEE Transactions on Geoscience and Remote Sensing}, 
  title={Tomographic Bistatic {3-D} Imaging and Coordinate Reconstruction Method Based on Uplink Communication Process}, 
  year={2025},
  volume={63},
  number={},
  pages={1-16},
  doi={10.1109/TGRS.2025.3556011}}

@ARTICLE{10636720,
  author={Lu, Bohao and Wei, Zhiqing and Wu, Huici and Zeng, Xinrui and Wang, Lin and Lu, Xi and Mei, Dongyang and Feng, Zhiyong},
  journal={IEEE Internet of Things Journal}, 
  title={Deep-Learning-Based Multinode {ISAC 4D} Environmental Reconstruction With Uplink–Downlink Cooperation}, 
  year={2024},
  volume={11},
  number={24},
  pages={39512-39526},
  doi={10.1109/JIOT.2024.3443648}}

@ARTICLE{10812728,
  author={Wen, Dingzhu and Zhou, Yong and Li, Xiaoyang and Shi, Yuanming and Huang, Kaibin and Letaief, Khaled B.},
  journal={IEEE Communications Surveys \& Tutorials}, 
  title={A Survey on Integrated Sensing, Communication, and Computation}, 
  year={2025},
  volume={27},
  number={5},
  pages={3058-3098},
  doi={10.1109/COMST.2024.3521498}}

@ARTICLE{10506595,
  author={Lyu, Zhidong and Zhang, Lu and Zhang, Hongqi and Yang, Zuomin and Zhang, Changming and Yang, Hang and Li, Nan and Bobrovs, Vjačeslavs and Ozolins, Oskars and Pang, Xiaodan and Liu, Guangyi and Yu, Xianbin},
  journal={Journal of Lightwave Technology}, 
  title={Multi-Channel Photonic {THz-ISAC} System Based on Integrated {LFM-QAM} Waveform}, 
  year={2024},
  volume={42},
  number={11},
  pages={3981-3988},
  doi={10.1109/JLT.2024.3392282}}

@ARTICLE{10872967,
  author={Zhang, Zhengyu and He, Ruisi and Ai, Bo and Yang, Mi and Zhang, Xuejian and Qi, Ziyi and Yuan, Yuan},
  journal={IEEE Transactions on Communications}, 
  title={Channel Measurements and Modeling for Dynamic Vehicular ISAC Scenarios at 28 {GHz}}, 
  year={2025},
  volume={73},
  number={8},
  pages={6884-6897},
  doi={10.1109/TCOMM.2025.3538851}}

@ARTICLE{11017419,
  author={Li, Lanxin and Liu, Che and Yu, Wenming and Cui, Tie Jun},
  journal={IEEE Sensors Journal}, 
  title={{Diff-HPE}: Multihypothesis Human Pose Estimation Based on {mmWave} Radar and Diffusion Framework}, 
  year={2025},
  volume={25},
  number={13},
  pages={25185-25197},
  doi={10.1109/JSEN.2025.3572253}}

@ARTICLE{11037950,
  author={Wang, Yingqi and Xu, Min and Wang, Zhongqin and Wang, Yongze and Zhang, J. Andrew},
  journal={IEEE Internet of Things Journal}, 
  title={User Reidentification Through {mmWave} Radio Imaging}, 
  year={2025},
  volume={12},
  number={16},
  pages={33293-33310},
  doi={10.1109/JIOT.2025.3576572}}

@ARTICLE{11028890,
  author={Chang, Yuance and Ding, Han and Cao, Feng and Zhao, Cui and Wang, Fei and Wang, Ge and Wang, Zhi and Xi, Wei},
  journal={IEEE Internet of Things Journal}, 
  title={{mmYodar+}: Robust Human Detection Using mmWave Signals}, 
  year={2025},
  volume={12},
  number={16},
  pages={33702-33713},
  doi={10.1109/JIOT.2025.3577559}}

@ARTICLE{11015573,
  author={Zhao, Langcheng and Lyu, Rui and Zhou, Anfu and Guo, Qi and Ma, Huadong},
  journal={IEEE Internet of Things Journal}, 
  title={{mmCG}: Noncontact Millimeter-Wave Cardiography for Heart Rate Variability Monitoring}, 
  year={2025},
  volume={12},
  number={15},
  pages={30955-30969},
  doi={10.1109/JIOT.2025.3573511}}

@ARTICLE{10770127,
  author={Niu, Yangyang and Wei, Zhiqing and Wang, Lin and Wu, Huici and Feng, Zhiyong},
  journal={IEEE Internet of Things Journal}, 
  title={Interference Management for Integrated Sensing and Communication Systems: A Survey}, 
  year={2025},
  volume={12},
  number={7},
  pages={8110-8134},
  doi={10.1109/JIOT.2024.3506162}}

@INPROCEEDINGS{10039194,
  author={Meng, Chunwei and Wei, Zhiqing and Feng, Zhiyong},
  booktitle={2022 14th International Conference on Wireless Communications and Signal Processing (WCSP)}, 
  title={Adaptive Waveform Optimization for MIMO Integrated Sensing and Communication Systems Based on Mutual Information}, 
  year={2022},
  volume={},
  number={},
  pages={472-477},
  doi={10.1109/WCSP55476.2022.10039194}}

@ARTICLE{11060842,
  author={Fang, Yin and Xu, Shu and Fang, Pan and Zhang, Jiexin and Huang, Yongming and Yang, Luxi},
  journal={IEEE Transactions on Vehicular Technology}, 
  title={Optimizing Codebook Design in mmWave Massive {MIMO} Using Digital Twin-Assisted {DRL}}, 
  year={2025},
  volume={},
  number={},
  pages={1-16},
  doi={10.1109/TVT.2025.3584798}}

@article{omondi2026advancing,
  title={Advancing MIMO adaptation with transfer learning: pioneering approaches and emerging perspectives},
  author={Omondi, Gevira and Olwal, Thomas O},
  journal={Cogent Engineering},
  volume={13},
  number={1},
  pages={2602250},
  year={2026},
  publisher={Taylor \& Francis}
}

@article{hu2024transfer,
  title={Transfer learning enabled transformer-based generative adversarial networks for modeling and generating terahertz channels},
  author={Hu, Zhengdong and Li, Yuanbo and Han, Chong},
  journal={Communications Engineering},
  volume={3},
  number={1},
  pages={153},
  year={2024},
  publisher={Nature Publishing Group UK London}
}

\begin{IEEEbiography}[{\includegraphics[width=1in,height=1.25in,clip,keepaspectratio]{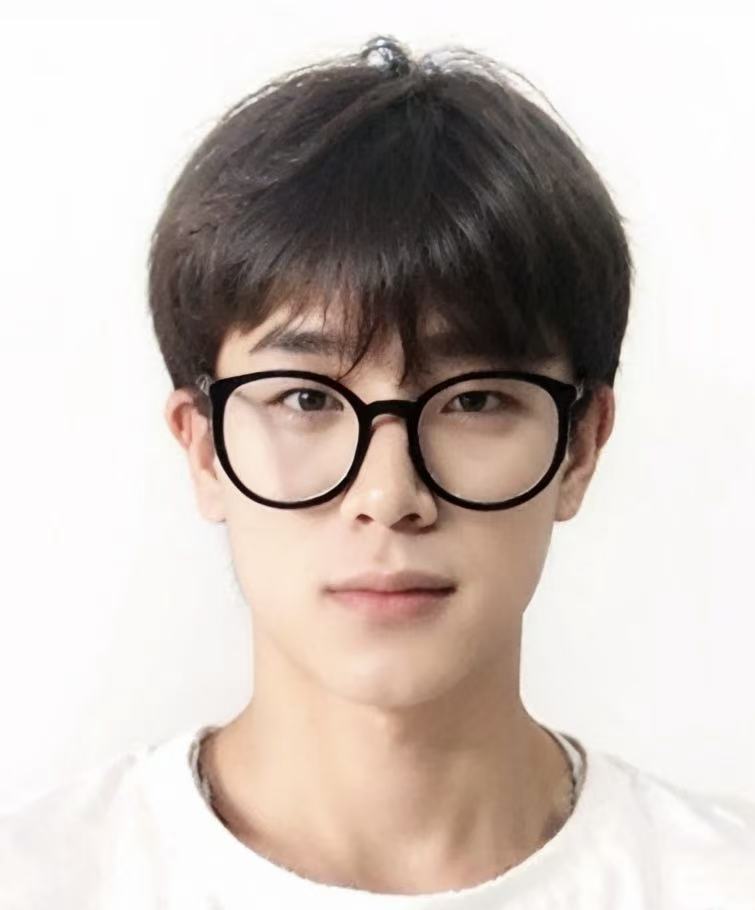}}]{Junzhe Song}
received the B.S. degrees in communication engineering from Nanjing University of Posts and Telecommunications, Nanjing, China, in 2023. He is currently pursuing the Ph.D. degree with the School of Electronic Information Engineering, Beijing Jiaotong University, Beijing, China. 

His research interests include ISAC, wireless communications, site-specific wireless channel modeling, and deep learning.
\end{IEEEbiography}
\vspace{11pt}
\begin{IEEEbiography}[{\includegraphics[width=1in,height=1.25in,clip,keepaspectratio]{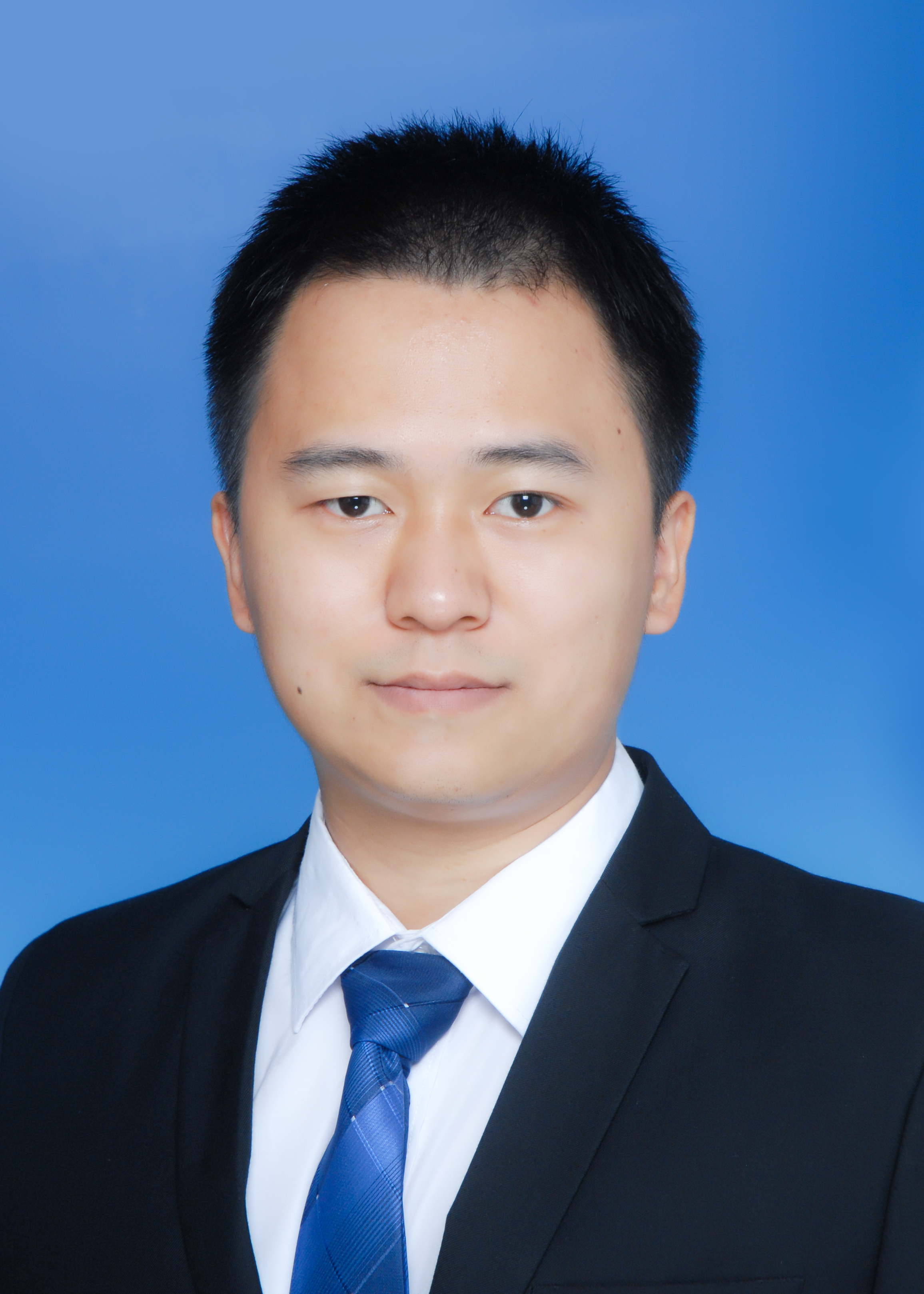}}]{Ruisi He}
(Senior Member, IEEE) received the B.E.
and Ph.D. degrees from Beijing Jiaotong University
(BJTU), Beijing, China, in 2009 and 2015, respec
tively.

He has been a Visiting Scholar with Georgia
Institute of Technology, Atlanta, GA, USA; the
University of Southern California, Los Angeles, CA,
USA; and the Universit\'e Catholique de Louvain,
Belgium. He is currently a Professor with the School
of Electronics and Information Engineering, BJTU.
He has authored/co-authored eight books, five book
chapters, more than 200 journal and conference papers, and several patents.
His research interests include wireless propagation channels and 5G and 6G
communications.

Dr. He served as the Early Career Representative (ECR) of Commission C,
International Union of Radio Science (URSI). He received the URSI Issac
Koga Gold Medal in 2021, the IEEE ComSoc Asia–Pacific Outstanding
Young Researcher Award in 2019, the URSI Young Scientist Award in
2015, and several Best Paper Awards in IEEE journals and conferences.
He has been an Editor of \textsc{IEEE Transactions on Communications}, 
\textsc{IEEE Transactions on Wireless Communications}, \textsc{IEEE Transac
tions on Antennas and Propagation}, \textit{IEEE Antennas and Propagation 
Magazine}, \textsc{IEEE Communications Letters}, \textsc{IEEE Open Journal of 
Vehicular Technology}, and a Lead Guest Editor of \textsc{IEEE Journal 
on Selected Area in Communications} and \textsc{IEEE Transactions on 
Antennas and Propagation}.
\end{IEEEbiography}
\vspace{11pt}
\begin{IEEEbiography}[{\includegraphics[width=1in,height=1.25in,clip,keepaspectratio]{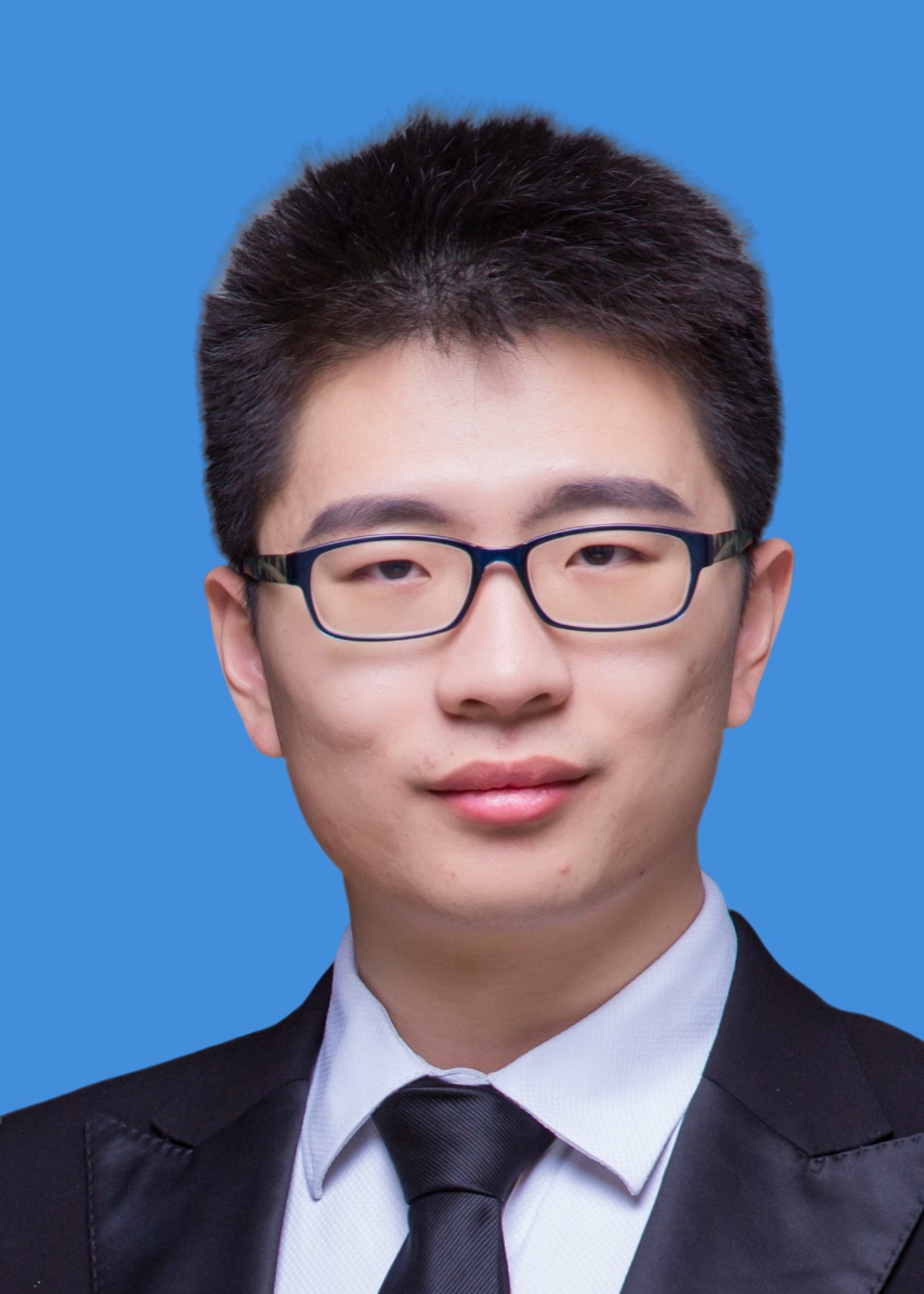}}]{Mi Yang}
(Member, IEEE) received the M.S. and
Ph.D. degrees from Beijing Jiaotong University,
Beijing, China, in 2017 and 2021, respectively.

He is currently an Associate Professor with the
School of Electronic and Information Engineering,
Beijing Jiaotong University. His research interests
are focused on wireless channel measurement and
modeling, vehicular and railway communications,
and AI-enabled channel research.
\end{IEEEbiography}
\vspace{11pt}
\begin{IEEEbiography}[{\includegraphics[width=1in,height=1.25in,clip,keepaspectratio]{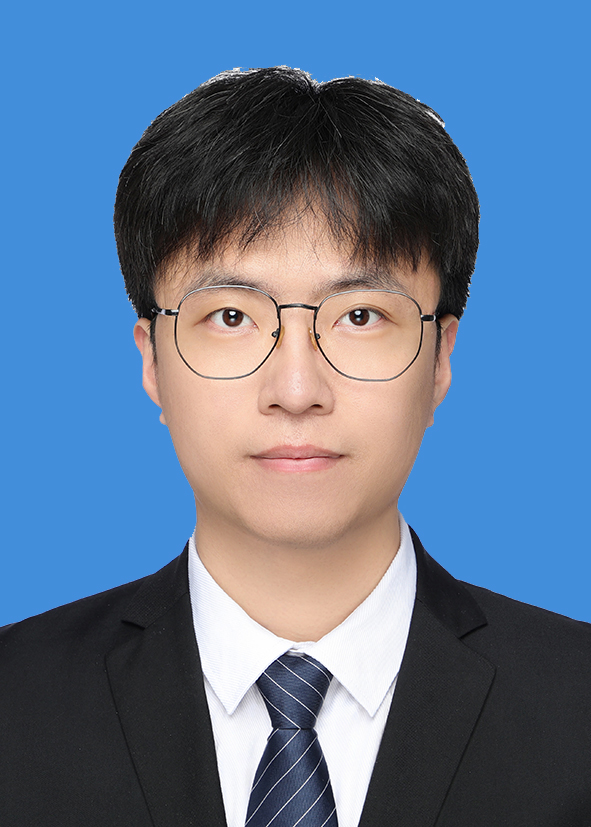}}]{Zhengyu Zhang}
(Student Member, IEEE) received
the B.S. and M.S. degrees in Electronic and Com
munication Engineering from Beijing Jiaotong Uni
versity, China, in 2018 and 2021, respectively, and
the Ph.D. degree in Information and Communication
Systems from Beijing Jiaotong University in 2025.
From 2024 to 2025, he was a visiting Ph.D. student
with the University of Bologna, Italy. 

He is currently a Lecturer with the School of Electronic and In
formation Engineering, Beijing Jiaotong University.
His research interests include integrated sensing and
communications (ISAC), channel semantics, deep learning, 6G wireless chan
nel measurement and modeling.
\end{IEEEbiography}
\vspace{11pt}
\begin{IEEEbiography}[{\includegraphics[width=1in,height=1.25in,clip,keepaspectratio]{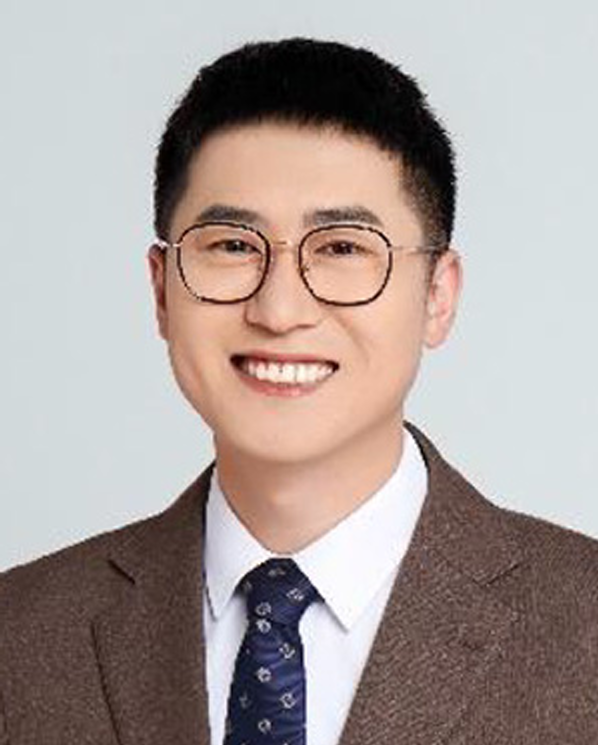}}]{Bingcheng Liu}
received the B.E. degree from
Beijing Jiaotong University, Beijing, China, in 2009,
and the M.S. degree from Beihang University,
Beijing, in 2012.

He is currently an Associate Professor with the
Aerospace Information Research Institute, Chinese
Academy of Sciences, Beijing. His work currently
focuses on the processing, analysis, and applica
tion of global navigation satellite system precision
localization and navigation.
\end{IEEEbiography}
\vspace{11pt}
\begin{IEEEbiography}[{\includegraphics[width=1in,height=1.25in,clip,keepaspectratio]{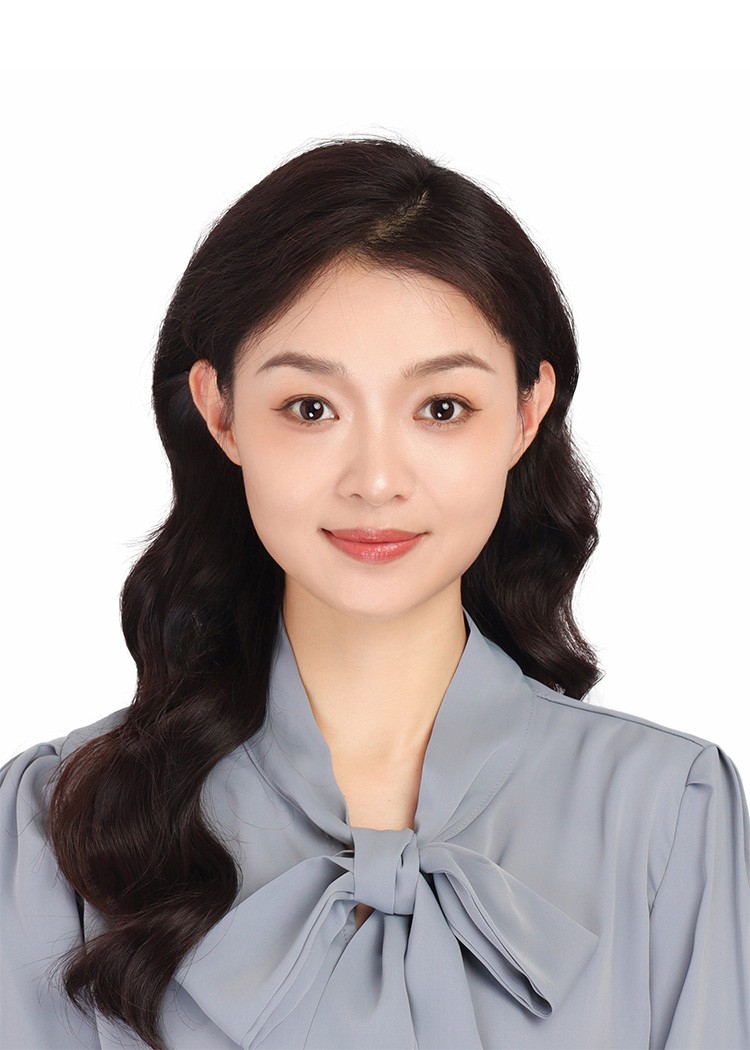}}]{Jiahui Han}
received the B.E. and M.S. degrees from
Beijing Jiaotong University, Beijing, China, in 2012
and 2016, respectively.

She is currently the Deputy Director of the
Scientific Research Management and Business
Development Department, Master Planning Insti
tute, China Academy of Industrial Internet, Beijing.
Her research focuses on information and com
munications, digital transformation for small- and
medium-sized enterprises, industrial internet, and
big data.
\end{IEEEbiography}
\vspace{11pt}
\begin{IEEEbiography}[{\includegraphics[width=1in,height=1.25in,clip,keepaspectratio]{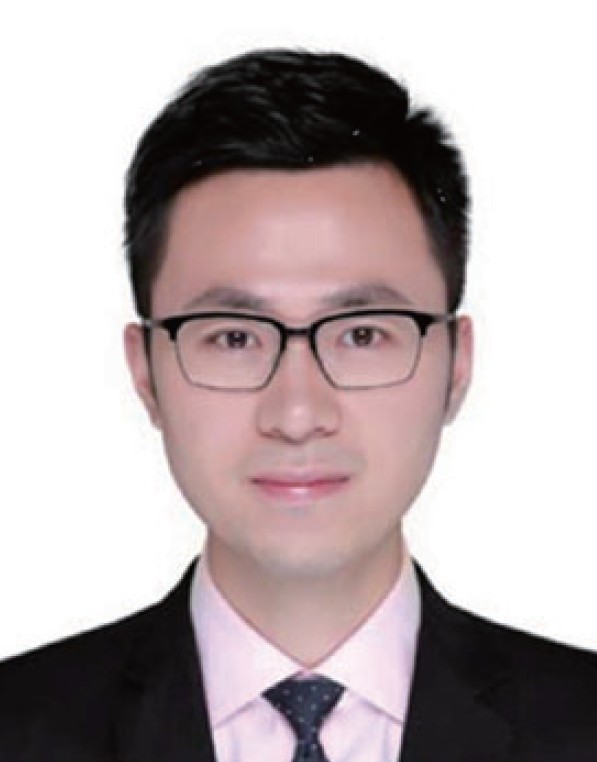}}]{Haoxiang Zhang}
received the B.E. and M.S.
degrees from Beijing Jiaotong University, Beijing,
China, in 2008 and 2012, respectively, and the Ph.D.
degree from Renmin University of China, Beijing, in
2019.

He is currently the Deputy Director (presiding)
of the Master Planning Institute, China Academy
of Industrial Internet, Beijing. His current research
interests include wireless communications, industrial
internet platform construction, regional planning,
industrial applications, and technical management.
Dr. Zhang is a member of the Editorial Board of \textit{Journal of New Industrialization.}
\end{IEEEbiography}
\vspace{11pt}
\begin{IEEEbiography}[{\includegraphics[width=1in,height=1.25in,clip,keepaspectratio]{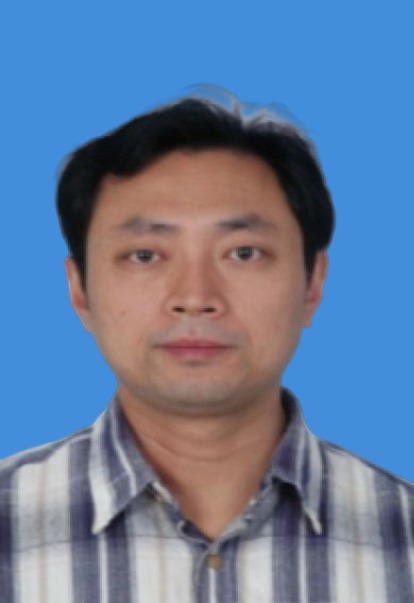}}]{Bo Ai}
(Fellow, IEEE) received the M.S. and Ph.D.
degrees from Xidian University, Xi’an, China, in
2002 and 2004, respectively.

He was with Tsinghua University, Beijing, China,
where he was an Excellent Post-Doctoral Research
Fellow in 2007. He is currently a Professor and an
Advisor of Ph.D. Candidates with Beijing Jiaotong
University, Beijing, where he is also the Deputy
Director of the State Key Laboratory of Rail Traffic
Control and Safety. He is also with the Engineer
ing College, Armed Police Force, Xi’an. He has
authored or co-authored six books and 270 scientific research papers, and
holds 26 invention patents in his research areas. His interests include the
research and applications of orthogonal frequency-division multiplexing tech
niques, high-power amplifier linearization techniques, radio propagation and
channel modeling, global systems for mobile communications for railway
systems, and long-term evolution for railway systems. He is a fellow of The
Institution of Engineering and Technology. He has received many awards, such
as the Qiushi Outstanding Youth Award by Hong Kong Qiushi Foundation, the
New Century Talents by Chinese Ministry of Education, the Zhan Tianyou
Railway Science and Technology Award by Chinese Ministry of Railways,
and the Science and Technology New Star by Beijing Municipal Science and
Technology Commission. He was the Co-Chair or the Session/Track Chair
for many international conferences, such as the 9th International Heavy Haul
Conference in 2009; the 2011 IEEE International Conference on Intelligent
Rail Transportation; HSRCom2011; the 2012 IEEE International Symposium
on Consumer Electronics; the 2013 International Conference on Wireless,
Mobile and Multimedia; IEEE Green HetNet 2013; and the IEEE 78th
Vehicular Technology Conference in 2014. He is an Associate Editor of IEEE
\textsc{TRANSACTIONS ON CONSUMER ELECTRONICS} and an Editorial Committee
Member of \textit{Wireless Personal Communications journal.}
\end{IEEEbiography}
\vspace{11pt}
\ifCLASSOPTIONcaptionsoff
  \newpage
\fi

 \end{document}